\documentclass[twocolumn]{aastex631}
\usepackage{graphicx}
\usepackage{xspace} 
\usepackage{enumerate}
\usepackage{multirow}
\usepackage{threeparttable}

\newcommand{\Msun}{\ensuremath{M_\odot}\xspace}
\newcommand{\OII}{[\textrm{O}~\textsc{ii}]\xspace}
\newcommand{\OIII}{[\textrm{O}~\textsc{iii}]\xspace}

\newcommand{\NII}{[\textrm{N}~\textsc{ii}]\xspace}
\newcommand{\SII}{[\textrm{S}~\textsc{ii}]\xspace}
\newcommand{\NeIII}{[\textrm{Ne}~\textsc{iii}]\xspace}
\begin{document}

\title{MAMMOTH-LyC: Investigating the Role of Galaxy Mergers in a Strong Lyman Continuum Leaker at $z=2.39$}

%%% Lead authors and program (admin) PIs

\author[0009-0007-6655-366X]{Shengzhe Wang}
\affiliation{National Astronomical Observatories, Chinese Academy of Sciences, Beijing 100101, China}
\affiliation{School of Astronomy and Space Science, University of Chinese Academy of Sciences (UCAS), Beijing 100049, China}

\author[0000-0002-9373-3865]{Xin Wang}
\affiliation{School of Astronomy and Space Science, University of Chinese Academy of Sciences (UCAS), Beijing 100049, China}
\affiliation{National Astronomical Observatories, Chinese Academy of Sciences, Beijing 100101, China}
\affiliation{Institute for Frontiers in Astronomy and Astrophysics, Beijing Normal University, Beijing 102206, China}

\author[0000-0001-6919-1237]{Matthew A. Malkan}
\affiliation{Department of Physics and Astronomy, University of California, Los Angeles, 430 Portola Plaza, Los Angeles, CA 90095, USA}

\author[0000-0002-7064-5424]{Harry I. Teplitz}
\affiliation{Infrared Processing and Analysis Center, Caltech, 1200 E. California Blvd., Pasadena, CA 91125, USA}

\author[0000-0002-3324-4824]{Rebecca L. Davies}
\affiliation{Centre for Astrophysics and Supercomputing, Swinburne University of Technology, Hawthorn, VIC 3122, Australia}

\author[0000-0002-3254-9044]{Karl Glazebrook}
\affiliation{Centre for Astrophysics and Supercomputing, Swinburne University of Technology, Hawthorn, VIC 3122, Australia}

%%% People performing analysis (mosfire reduction, galfit, photometric catalogs, etc.)

\author[0000-0001-6505-0293]{Keunho J. Kim}
\affiliation{Infrared Processing and Analysis Center, Caltech, 1200 E. California Blvd., Pasadena, CA 91125, USA}

\author[0000-0003-2804-0648]{Themiya Nanayakkara}
\affiliation{Centre for Astrophysics and Supercomputing, Swinburne University of Technology, Hawthorn, VIC 3122, Australia}

\author[0009-0004-7133-9375]{Hang Zhou}
\affil{School of Astronomy and Space Science, University of Chinese Academy of Sciences (UCAS), Beijing 100049, China}

\author[0000-0002-0663-814X]{Yiming Yang}
\affil{National Astronomical Observatories, Chinese Academy of Sciences, Beijing 100101, China}
\affil{School of Astronomy and Space Science, University of Chinese Academy of Sciences (UCAS), Beijing 100049, China}

\author[0000-0002-9390-9672]{Chao-Wei Tsai}
\affil{National Astronomical Observatories, Chinese Academy of Sciences, Beijing 100101, China}
\affil{Institute for Frontiers in Astronomy and Astrophysics, Beijing Normal University,  Beijing 102206, China}
\affil{School of Astronomy and Space Science, University of Chinese Academy of Sciences (UCAS), Beijing 100049, China}

\author[0009-0005-3823-9302]{Yuxuan Pang}
\affiliation{School of Astronomy and Space Science, University of Chinese Academy of Sciences (UCAS), Beijing 100049, China}

\author[0000-0001-8467-6478]{Zheng Cai}
\affiliation{Department of Astronomy, Tsinghua University, Beijing 100084, China}

\author[0000-0003-3310-0131]{Xiaohui Fan}
\affiliation{Steward Observatory, University of Arizona, 933 North Cherry Ave., Tucson, AZ 85721, USA}

\author[0000-0002-6586-4446]{Alaina Henry}
\affiliation{Space Telescope Science Institute, 3700 San Martin Drive, Baltimore, MD 21218, USA}

\author[0000-0001-5951-459X]{Zihao Li}
\affil{Cosmic Dawn Center (DAWN), Denmark}
\affil{Niels Bohr Institute, University of Copenhagen, Jagtvej 128, DK2200 Copenhagen N, Denmark}

\author[0000-0002-3264-819X]{Dong Dong Shi}
\affiliation{Center for Fundamental Physics, School of Mechanics $\&$ Optoelectronic Physics, Anhui University of Science and Technology, Huainan 232001, China}

\author[0000-0003-3728-9912]{Xian~Zhong Zheng}
\affiliation{Tsung-Dao Lee Institute and State Key Laboratory of Dark Matter Physics, Shanghai Jiao Tong University, Shanghai 201210, China}

\author[0009-0002-7871-3337]{Zhiyu Yan}
\affil{National Astronomical Observatories, Chinese Academy of Sciences, Beijing 100101, China}

\correspondingauthor{Xin Wang}
\email{xwang@ucas.ac.cn}

\begin{abstract}

The MAMMOTH-LyC survey is a cycle 30 Hubble Space Telescope (HST) medium program obtaining 18-orbit-deep WFC3/UVIS F225W imaging in two massive galaxy protocluster fields at $z\sim2.2$.
We introduce this survey by reporting the discovery of J1244-LyC1, a strong Lyman continuum (LyC) leaker at $z = 2.39$, exhibiting clear merger signatures. J1244-LyC1 has a highly significant ($6\sigma$) LyC detection, corresponding to an absolute escape fraction of $f_{\mathrm{esc}} \! = 36\%\pm6\%$ ($1\sigma$). The LyC emission is spatially resolved into multiple peaks that coincide with the system's disturbed morphology, confirming genuine multi-site LyC leakage. With a stellar mass of $10^{10.2}\Msun$, J1244-LyC1 is both the first confirmed high-redshift LyC-leaking merger and the most massive LyC emitter known to date. We interpret J1244-LyC1 as a merger-driven starburst system in which tidal interactions have disrupted the interstellar medium, creating multiple low-column-density pathways that facilitate LyC escape. This discovery provides the first direct evidence of spatially resolved LyC escape in a merging system, offering new insight into the potential role of major mergers in driving the cosmic reionization.

\end{abstract}


\section{Introduction} \label{sect:intro}

% Lyman continuum (LyC) photons ($\lambda_{\rm rest}<912$\,\AA) produced by galaxies powered the last major cosmic phase transition---the Epoch of Reionization \citep[EoR, $z\sim6$--11;][]{stark_galaxies_2016}, during which the neutral intergalactic medium (IGM) became reionized \citep{Dayal:2020ki}. However, the fraction of ionizing photons that successfully escape their host galaxies, $f_{\mathrm{esc}}$, remains one of the most crucial unknowns in understanding the EoR. Constraining this parameter directly addresses a central and long-standing question: what sources reionized the Universe? \citep{finkelstein_conditions_2019,naidu_rapid_2020}.

Lyman continuum (LyC) photons ($\lambda_{\rm rest}<912$\,\AA) produced by the first luminous sources drove the last major cosmic phase transition---the Epoch of Reionization \citep[EoR, $z\sim6$--11;][]{stark_galaxies_2016}, during which the predominantly neutral intergalactic medium (IGM) became ionized \citep{Dayal:2020ki}. Star-forming galaxies are generally thought to have supplied most of the required ionizing photons, but the dominant contributing population remains uncertain: the ionizing budget may have been provided by numerous faint, low-mass galaxies \citep[e.g.,][]{finkelstein_conditions_2019}, by less numerous but more UV-luminous galaxies with high escape fractions \citep[e.g.,][]{naidu_rapid_2020}, or, at least in part, by active galactic nuclei (AGNs; \citealt{Jiang_2025}). Distinguishing among these scenarios requires determining how the escape fraction of ionizing photons, $f_{\mathrm{esc}}$, varies with galaxy luminosity, stellar mass, and other physical properties.

Directly observing LyC-leaking galaxies is essential for answering this question, yet such efforts face a fundamental challenge: the rapidly declining IGM transmission for LyC photons at $z>4$ \citep{inoue_updated_2014}, which renders direct detections infeasible at the redshifts most relevant to reionization. As a result, observational searches for LyC emission have focused on two lower-redshift windows: systems at $z\sim0.3-0.5$ accessible with the Cosmic Origins Spectrograph (COS) on the Hubble Space Telescope (HST) \citep[e.g.,][]{Izotov_2016,Malkan2021,wang__low-redshift_2021}, and galaxies at $z\sim2$--4—the “cosmic noon” epoch—using HST/WFC3-UVIS imaging \citep[e.g.,][]{Jones_2021,Kerutt_2024,marques-chaves_witnessing_2024,wang_lyman_2025,beckett2025parallelionizingemissivitysurvey}.

At the low-redshift window ($z\sim0.3$), HST/COS spectroscopy, particularly through the Low-redshift Lyman Continuum Survey (LzLCS), has generated a sample of confirmed LyC leakers \citep{Izotov_2016,Flury_2022_II,Malkan2021}.
These samples have revealed correlations between $f_{\mathrm{esc}}$ and several observable diagnostics, including the UV continuum slope~$\beta$ \citep{Chisholm_2022_UV_slope}, the Ly$\alpha$ emission-line profile \citep{Flury_2022_II,naidu_synchrony_2022}, and the \OIII/\OII\ ratio (O32) \citep{Pellegrini_2012, Li2018}. 
As summarized by \citet{Jaskot_lowz_review_2025}, although massive LyC leakers have also been reported at low redshift \citep{Roy_2025}, existing statistical samples are dominated by lower-mass systems. Confirmed LyC leakers tend to have higher specific star formation rates (sSFR; \citealt{Leitet_2013}), extremely high ionization parameters, and low H\,\textsc{i} covering fractions along the line of sight.
This implies that the ISM is highly disrupted and porous, with anisotropy that is associated with extreme star-formation activity \citep{carr_2025,Jaskot_lowz_review_2025}.
Consequently, powerful stellar feedback processes are often invoked to explain the observed correlations \citep{Jaskot_2017,Komarova_2021_MK71,Flury_2025,carr_2025,Gazagnes_2018,Flury_2022_II}. 
However, the specific physical mechanisms that simultaneously trigger starbursts and enable LyC escape remain poorly understood.

% Observationally, a key piece of evidence at low redshift comes from the Ly$\alpha$ and Continuum Origins Survey (LaCOS; \citealt{Le_Reste_2025_I}), an HST imaging follow-up of 42 confirmed LyC leakers drawn from the LzLCS+ sample. A subsequent morphological analysis found that more than 41\% of these systems exhibit merger signatures, suggesting that galaxy interactions may represent an important channel for facilitating LyC escape \citep{Le_Reste_2025_III}.

Galaxy mergers have long been proposed as a physical mechanism connected to LyC escape \citep{Mascia_2025}. 
Observationally, a key piece of evidence at low redshift comes from the Ly$\alpha$ and Continuum Origins Survey (LaCOS; \citet{Le_Reste_2025_I}),an HST imaging follow-up of 42 confirmed LyC leakers drawn from the LzLCS+ sample. A subsequent morphological analysis found that more than 41\% of these systems exhibit merger signatures, suggesting that galaxy interactions may be an important facilitator of LyC escape \citep{Le_Reste_2025_III}. 
In addition, the nearest major merger system, the ''Antennae'' Galaxy (NGC~4038/39; \citealt{Weilbacher_2018}), also exhibits signatures of ionizing photon escape associated with its tidal structures.
In theory, gas-rich galaxy mergers can promote LyC escape through two primary pathways. 
First, mergers can trigger intense star formation while simultaneously disturbing the gas distribution, creating favorable conditions for LyC escape \citep{Le_Reste_2025_III,Le_Reste_2024,yuan_merging_2024,zhu_lyman_2024,Faria_2025}. 
Second, tidal forces can expel gas from galactic centers, forming substructures such as tidal tails where star formation can occur \citep{Pearson_2016}. 
Through the depletion of gas reservoirs in these tidal features and the effects of stellar feedback, low-column-density channels may form, facilitating the escape of ionizing photons. 
Recent simulations by \citet{Ejdetj_2026} have demonstrated that dwarf galaxy mergers can reproduce the observed properties of the nearby LyC leaker Haro~11 \citep{Komarova_2024}. 
Consistently, cosmological simulations of the EoR also suggest that galaxy mergers play an important role in enhancing LyC escape \citep{kostyuk_influence_2024}.

% Galaxy mergers have long been proposed as a mechanism capable of enhancing LyC escape by reshaping the gas morphology and star-formation activity \citep{Bridge_2010,Purkayastha_2022_GPs,Le_Reste_2024,yuan_merging_2024,zhu_lyman_2024}. Gas-rich interactions frequently trigger repeated starbursts after pericentric passages \citep{Faria_2025}, and merging systems tend to exhibit significantly elevated star-formation rates (SFRs) relative to isolated galaxies, especially during the late stages of interaction \citep{Patton_2013_galaxy_pair,Stierwalt_2015,Ferreira_2025}. These intense starbursts efficiently produce large populations of massive stars---thereby increasing the intrinsic LyC photon budget---and can facilitate LyC escape through stellar and supernova-driven feedback \citep{Trebitsch_2017,Barrow_2020,Ma_2020_lyc,Choustikov_2024}. In addition, tidal forces can redistribute gas away from galactic centers \citep{Pearson_2016}, potentially boosting $f_{\mathrm{esc}}$ along lines of sight that pass outside the tidally displaced material \citep{Le_Reste_2024,Ejdetj_rn_2025}. Cosmological simulations at $z\sim5$--10 likewise find that mergers can enhance LyC leakage under simplified $f_{\mathrm{esc}}$ prescriptions \citep{kostyuk_influence_2024}. Observationally, in the low-$z$ regime, the Ly$\alpha$ and Continuum Origins Survey (LaCOS) has compiled a sample of 42 LyC leakers, confirming that more than 41\% reside in merging systems and establishing mergers as an important channel for LyC escape \citep{Le_Reste_2025_III}.

In contrast, within the high-redshift window ($z\sim2$--4), although many LyC candidates show clear morphological signatures of galaxy interactions, spatial offsets between LyC-band emission and the UV continuum have made it extremely difficult to rule out contamination from low-$z$ interlopers \citep{zhu_lyman_2024}. For example, Ion3 \citep{Me_tri__2025,Vanzella_2018} exhibits strong ground-based spectroscopic features yet is dominated by a foreground interloper; its true LyC detection significance is only $3.5\sigma$. Other candidates such as z19863 and CDFS-6664 \citep{gupta_mosel_2024,yuan_merging_2024} show similar spatial offsets. 
% The intrinsically weak UV continuum associated with these LyC-leaking regions further complicates interpretation and forces a re-evaluation of the physical origin of the claimed LyC signals.
% In principle, unambiguous confirmation requires high spatial-resolution spectroscopy. Consequently, no high-redshift merger system has yet been securely demonstrated to exhibit significantly enhanced $f_{\mathrm{esc}}$.
In cases where the LyC signal is spatially offset from the main body of the merging system, the authenticity of the detected LyC emission must be carefully assessed. In principle, an unambiguous confirmation requires high spatial-resolution spectroscopy \citep{Rivera-Thorsen_2025}. Although contamination from low-redshift interlopers cannot be completely ruled out, the statistical prevalence of such systems nevertheless indicates that merging galaxies constitute a substantial fraction of the high-redshift LyC leaker population.

In this paper, we report the discovery of a new LyC emitter at $z=2.39$ in the BOSS1244 field. This source benefits from extensive multi-band imaging and spectroscopy from both HST and Keck, enabling a detailed investigation of its physical properties. The structure of this paper is as follows: Section~\ref{sect:observation} describes the observations and data reduction; Section~\ref{sect: analysis} presents our analysis, including spectral-line modeling, spectral energy distribution (SED) fitting, and the calculation of $f_{\mathrm{esc}}$; and Section~\ref{sect: discussion} discusses the implications of our results. Throughout this work, we adopt a flat $\Lambda$CDM cosmology with $H_0 = 70~\mathrm{km~s^{-1}~Mpc^{-1}}$, $\Omega_{\mathrm{M}} = 0.3$, and $\Omega_{\Lambda} = 0.7$.

\section{Observations} \label{sect:observation}

The primary dataset for J1244-LyC1 is drawn from the MAMMOTH (MApping the Most Massive Overdensity Through Hydrogen) program series \citep{Cai_Zheng_2016,Cai_Zheng_2017a,Xin_Wang_2022_MAMMOTH, Zhou_hang_2025, Yiming_2025,Emmet_2025}. These programs target the BOSS1244 protocluster at $z = 2.24 \pm 0.02$, one of the most massive known overdensities at cosmic noon \citep{Cai_Zheng_2016, Cai_Zheng_2017a}. The HST Cycle-28 medium program, the MAMMOTH-Grism survey (GO-16276; P.I.: X.~Wang; \citealt{Xin_Wang_2022_MAMMOTH}), provided spectroscopic identification across the overdensity, detecting key rest-frame optical lines such as [O\,{\sc iii}], [O\,{\sc ii}], and H$\beta$.

\begin{figure*}[htbp]
    \centering
    \includegraphics[width=1\textwidth]{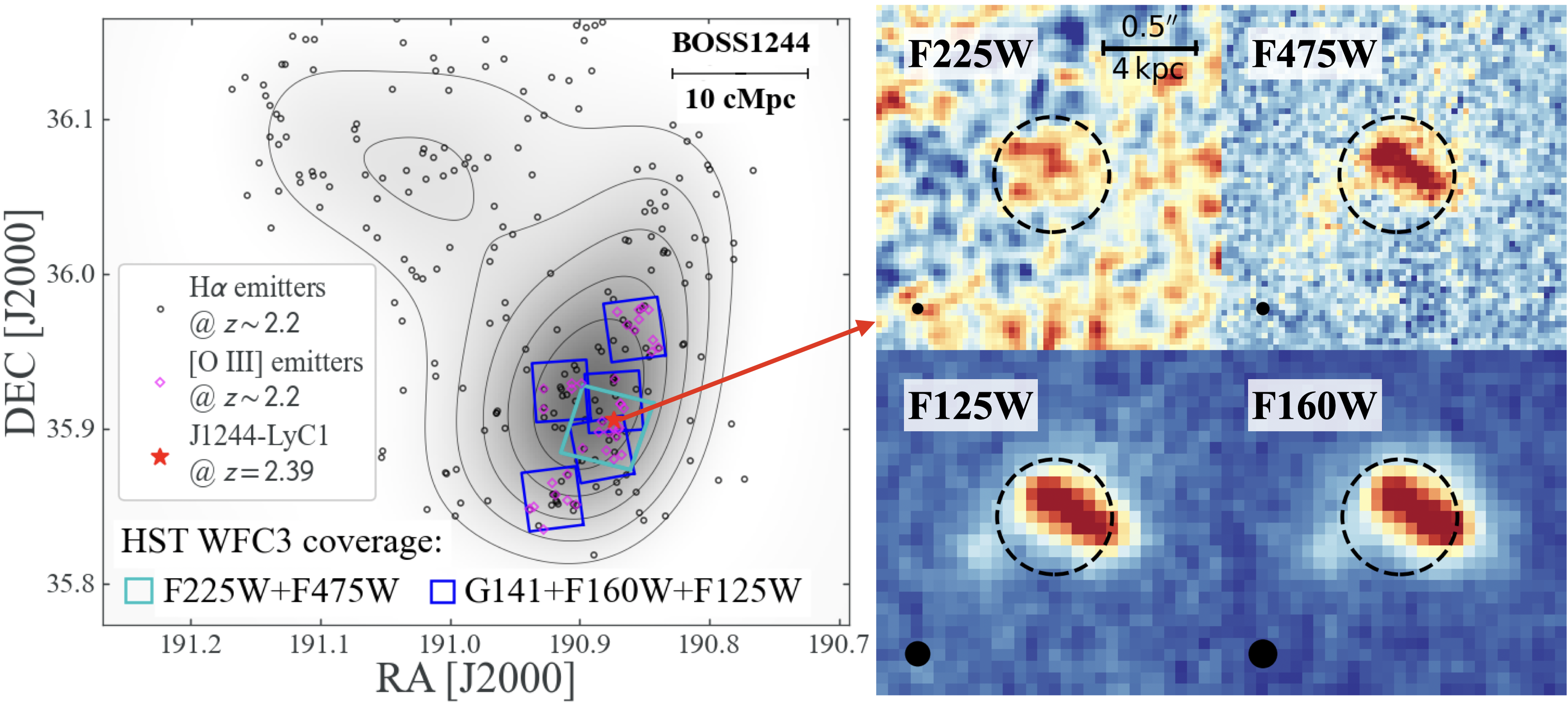}
    \caption{
    HST WFC3 coverage of the BOSS1244 protocluster field and high-resolution multi-band imaging of J1244-LyC1. 
    {\sc Left:} The MAMMOTH-Grism and MAMMOTH-LyC programs target overdense fields of extreme emission-line galaxies (EELGs) at $z \approx 2.2$, corresponding to the BOSS1244 protocluster. 
    Black circles mark spectroscopically confirmed H$\alpha$ emitters (HAEs; \citealt{Shi_2021}), while the magenta diamonds indicate EELGs with ${\rm EW}$(\OIII)$ > 225\,\text{\AA}$ following \citet{Tang_2019_OIII}, identified using the MAMMOTH-Grism deep HST grism spectroscopy. 
    The red star denotes the location of J1244-LyC1. 
    {\sc Right:} HST imaging of J1244-LyC1 in WFC3/UVIS F225W (rest-frame LyC; PSF-smoothed), F475W (rest-frame UV), and WFC3/IR F125W and F160W (rest-frame optical). 
    The FWHM of the PSF is shown by the black circle. 
    The 0.6\arcsec aperture is indicated by the black dashed circle.
    Each cutout is 1.8\arcsec \,$\times$\,1.8\arcsec in size.
    % \vspace{-0.5em}
    }\label{fig:1} 
\end{figure*}

In addition, the HST Cycle-30 medium program MAMMOTH-LyC (GO-17159; P.I.: X.~Wang) obtained 18 orbits of ultra-deep LyC-band (F225W) imaging—probing LyC emission at $z\!>\!2.2$—over the BOSS1244 field (Fig.~\ref{fig:1}). 
% The HST observations are complemented by extensive ground-based multi-band data, including CFHT $Ks$ band, LBT $U$ and $z$ band imaging, and Keck/MOSFIRE $K$ band spectroscopy covering H$\alpha$ \citep{Zhou_hang_2025}. 
The HST observations are complemented by extensive ground-based multi-band data, including $K_s$-band imaging obtained with the Wide-field InfraRed Camera (WIRCam) on the Canada-France-Hawaii Telescope (CFHT), $U$- and $z$-band imaging obtained with the Large Binocular Camera (LBC) on the Large Binocular Telescope (LBT), and Keck/MOSFIRE $K$-band spectroscopy covering H$\alpha$ \citep{Zhou_hang_2025}.

\subsection{HST Data and Reduction} \label{subsect: data}

High-resolution HST imaging of J1244-LyC1 was obtained with WFC3/UVIS and WFC3/IR from several programs (GO-17159 and GO-16276, P.I.: X.~Wang; \citealt{Xin_Wang_2022_MAMMOTH}; GO-15266, P.I.: Z.~Cai; \citealt{Liu_Shuang_2023}). J1244-LyC1 was imaged in the F225W, F475W, F125W, and F160W filters, with total exposure times of $\sim$48,600\,s, $\sim$2800\,s, $\sim$1800\,s, and $\sim$2600\,s, respectively. These probe rest-frame LyC (600–800\,\AA), UV (1400\,\AA), and optical (3700–4750\,\AA) emission. Slitless spectroscopy from the MAMMOTH-Grism survey provides WFC3/G141 grism data ($R \sim 100$), covering \OII and H$\beta$ for J1244-LyC1 \citep{Xin_Wang_2022_MAMMOTH}.

% Ground-based imaging includes LBT/LBC $U$- and $z$-band observations and CFHT/WIRCam $K$-band data, with total exposure times of 4.7\,hr, 4\,hr, and 5\,hr, respectively. 
% The observations were taken under seeing conditions of 0.8\arcsec–1\arcsec. 

% \subsection{WFC3/UVIS Data Reduction} \label{subsect: UVIS reduction}

% We used the pipeline-calibrated FLC (xxxspell that out XXX)single exposures from MAST as the starting point for UVIS reduction. 
We used the pipeline-calibrated, CTE-corrected {\tt \_flc.fits} single exposures from MAST as the starting point for the UVIS reduction.
Cosmic rays were identified and masked using {\tt astroscrappy}. To correct amplifier-dependent background variations, we equalized the background levels across the four readout amplifiers.

Multiple exposures were aligned using an iterative astrometric refinement procedure. The F475W mosaic was first registered to the F160W reference frame from the MAMMOTH-Grism survey, and the F225W exposures were subsequently aligned to the F475W frame. Final mosaics were produced using {\tt AstroDrizzle} v3.7.1 from {\tt DrizzlePac} \citep{Fruchter_2002}, following the configuration described in \citet{carter2025skysurf10novelmethodmeasuring} .
% (details in Appendix~\ref{sect: app_B}).
Our input parameters for {\tt Astrodrizzle} are listed in Appendix~\ref{sect: app_B}.
The pixel size of the final F225W and F475W mosaics is 0.03\arcsec.

The most significant challenge arose from aligning the ultra-deep 18-orbit F225W imaging. The UV field contains few suitable stars for astrometric solutions, and roughly 30\% of the exposures are affected by cosmic-ray contamination, due to full orbit long exposures. In addition, increased pointing uncertainties due to aging HST gyroscopes further complicated the alignment.

To address these issues, we developed a customized reduction framework incorporating affine transformations, iterative drizzling, cosmic-ray rejection, refined photometric matching, and density-based clustering. 
This procedure achieved a final relative astrometric precision of $\sim$0.2\,pixel, i.e., $\sim$6 milli-arcsec.
% The method is summari\zed in Appendix~\ref{sect: app_A}, with a full description to be presented in a forthcoming paper (Wang et al., in preparation). 
A detailed description of the method will be presented in the forthcoming data release and initial science results (Wang et al., in preparation).

\subsection{Keck Observations} \label{subsect: Keck}

J1244-LyC1 was observed twice using Keck/MOSFIRE:  
2022A\_U016 (P.I.: M.~Malkan; \citealt{Zhou_hang_2025}) and  
2025A\_W335 (P.I.: R. Davies).
% The first observation was obtained on 15 April 2022 in Multi-Object Spectroscopy (MOS) mode, with a total exposure time of 7920\,s (11 $\times$ 4 $\times$ 180\,s) under $\sim$0.7\arcsec\ seeing. The second observation, carried out on 21 February 2025 in Long-Slit Spectroscopy (LSS) mode, accumulated 3960\,s (22 $\times$ 180\,s) under $\sim$0.6\arcsec\ seeing.
The first observation was obtained on 15 April 2022 in Multi-Object Spectroscopy (MOS) mode, with a 0.7\arcsec\ slit width and a total exposure time of 7920\,s (11 $\times$ 4 $\times$ 180\,s) under $\sim$0.7\arcsec\ seeing. 
The second observation, carried out on 21 February 2025 in Long-Slit Spectroscopy (LSS) mode, used a 1.0\arcsec\ slit width and accumulated 3960\,s (22 $\times$ 180\,s) under $\sim$0.6\arcsec\ seeing.
The first observation only covered part of the source, the second observation was at a different PA that allowed it to cover both main components.

The reduction procedure follows \cite{Zhou_hang_2025}. MOSFIRE data were processed with the {\tt PYPEIT} pipeline \citep{Prochaska_pypeit_2020}, which performs wavelength calibration using atmospheric OH emission lines. One-dimensional spectra from individual exposures were extracted and co-added to increase the signal-to-noise ratio (SNR). Additional details of our reduction steps are provided in Appendix~\ref{sect: app_C}.

\subsection{Ground-based Imaging} \label{subsect: Ground-based}
Ground-based imaging includes LBT/LBC  
% xxx spell that out the first time XXX 
$U$  and $z$ band observations and CFHT/WIRCam $Ks$ band data, with total exposure times of 4.7\,hr, 4\,hr, and 5\,hr, respectively. 
The observations were taken under seeing conditions of 0.8\arcsec–1\arcsec. 
These ground-based photometric measurements provide essential constraints for the global SED fitting of J1244-LyC1.

%%%%%%%%%%%%%%%%%%%%%%%%%%%%%%%%%%%%%%%%%%%%%%%%%%%%%%%%%%%%%
\section{ANALYSIS} \label{sect: analysis}

\subsection{LyC Detection} \label{subsect: LyC_D}

Fig.~\ref{fig:1} presents the high-spatial-resolution HST imaging of J1244-LyC1 in the F225W, F475W, F125W, and F160W filters.  
We adopt the F160W image as the detection reference and use its $3\sigma$ segmentation map to define a common photometric aperture.  
% The flux in each PSF-matched image was measured by integrating all pixels within this F160W-defined aperture. 
% Because the drizzling process introduces correlated noise that is not fully captured by the inverse-variance weight maps,
% we calibrated the propagated photometric uncertainty empirically using randomly placed apertures in nearby source-free regions, adopting the aperture-dependent correction appropriate for the circularized size of the target aperture.
% The flux in each PSF-matched image was measured by integrating all pixels within the F160W-defined aperture. 
The flux in each PSF-matched image was measured within the F160W-defined aperture and corrected for aperture losses.
Because image resampling introduces pixel-to-pixel correlations that are not fully captured by independent-pixel error propagation \citep{Casertano_2000}, we empirically calibrated the F225W photometric uncertainties derived from the {\tt AstroDrizzle} inverse-variance weight (WHT) map. Following the empty-aperture approach of \citet{Labbe_2003} and \citet{Gawiser_2006}, we placed random circular apertures over a range of radii in nearby source-free regions and measured the standard deviation of their enclosed fluxes, $\sigma_{\rm sky}(r)$. We defined the aperture-dependent correction factor as $C_{\rm corr}(r)=\sigma_{\rm sky}(r)/\sigma_{\rm WHT}(r)$, where $\sigma_{\rm WHT}(r)$ is the corresponding aperture uncertainty predicted from the WHT map.

For J1244-LyC1, the F160W-defined source aperture has a circularized radius of approximately 8 pixels ($0.24\arcsec$ at the F225W pixel scale of $0.03\arcsec\,{\rm pixel}^{-1}$). At this aperture size, we measure $C_{\rm corr}\simeq2.7$ and therefore adopt $\sigma_{\rm F225W}=C_{\rm corr}\sigma_{\rm WHT,src}\simeq2.7\,\sigma_{\rm WHT,src}$. This empirical correction accounts for the excess aperture noise from pixel correlations and residual background fluctuations. We use an aperture-dependent correction because $C_{\rm corr}$ continues to increase with aperture radius over the range examined. The corrected uncertainty yields a total F225W detection significance of approximately $6\sigma$. All quoted F225W photometric uncertainties and detection significances incorporate the correction appropriate to their aperture size.
Within this region, the PSF-matched LyC-band measurement yields a $6\sigma$ detection, with $m_{\rm F225W}=27.81^{+0.20}_{-0.17}\,\mathrm{AB~mag}$ (Table~\ref{tab:measure}).

% Figure~\ref{fig:1} shows the spatial distribution of the LyC emission relative to the rest-frame UV morphology.  
The LyC emission displays an extended structure whose centroid does not coincide with either of the two UV luminosity peaks.  
We further analyze this behavior in Section~\ref{subsect: S_fesc_c}, where the system is decomposed into three principal LyC-leaking regions.  
Although the apparent offset between LyC emission and UV peaks resembles that seen in z19863 and CDFS-6664 \citep{gupta_mosel_2024, yuan_merging_2024}, the dominant LyC-emitting region in J1244-LyC1 lies near the central interface of the merging system, distinct from previously reported cases.

\subsection{Spectrum Analysis} \label{subsect: Spec_A}

\subsubsection{Emission Lines} \label{subsubsect: EL}

J1244-LyC1 is covered by both \textit{HST} WFC3/G141 grism data ($R\sim100$) and two Keck/MOSFIRE $K$ band observations ($R=3600$; Appendix~\ref{sect: app_D}).  
The spectra detect \OII, H$\beta$, and H$\alpha$ line emission, with no additional strong emission features.  
These lines confirm a redshift of $z=2.39$, and no evidence is found for a low-$z$ interloper.

A portion of the \OIII$\lambda4959$ line is marginally detected at the edge of the G141 grism coverage. However, because the throughput declines sharply at the bandpass edge and the flux calibration is unreliable, we do not attempt to estimate \OIII fluxes.

Slit-loss corrections (typically a factor of 1.2--2 
% xxx or Whatever XXX
)  were applied to both MOSFIRE datasets by convolving the F160W image with the seeing of each night and comparing the resulting flux distribution.  
The median seeing was $0.7\arcsec$ for the first observation and $0.6\arcsec$ for the second.

Emission-line fluxes were then measured from the G141 grism and MOSFIRE spectra (see Table~\ref{tab:measure}).  
No significant \NII, \SII or \NeIII detections are found in the spectra. 
% and \NeIII $3\sigma$ upper limits are reported in Table~\ref{tab:measure}.

\begin{deluxetable*}{lccccccccc}
\tablecaption{Emission line and photometry of J1244-LyC1 \label{tab:measure}}
\tablehead{
\colhead{H$\alpha$} & 
\colhead{H$\beta$} & 
\colhead{\OII} & 
% \colhead{\NeIII \tablenotemark{a}} & 
\colhead{HST F225W} & 
\colhead{LBT $U$} & 
\colhead{HST F475W} & 
\colhead{LBT $Z$} & 
\colhead{HST F125W} &
\colhead{HST F160W} &
\colhead{CFHT $Ks$} 
}
\startdata
$8.4\pm0.4$  & $2.6\pm0.6$ &  $5.6\pm0.7$ & 
% $<2$ & 
$27.81^{+0.20}_{-0.17}$ & $25.15^{+0.10}_{-0.09}$ & $24.27^{+0.01}_{-0.01}$ &
$24.14^{+0.09}_{-0.08}$ & $23.57^{+0.01}_{-0.01}$ &
$23.26^{+0.01}_{-0.01}$ & $22.80^{+0.18}_{-0.17}$ \\
\enddata
\tablecomments{
Observed line fluxes are measured from the WFC3/G141 grism data and Keck MOSFIRE spectra. 
The unit of line fluxes and multi-band photometry is 
$10^{-17}~\mathrm{erg\,s^{-1}\,cm^{-2}}$ and AB~mag.
All reported uncertainties represent 1$\sigma$ errors.
}
% \tablenotetext{a}{3$\sigma$ upper limit.}
\end{deluxetable*}

\begin{figure*}[htbp]
    \centering
    \includegraphics[width=1\textwidth]{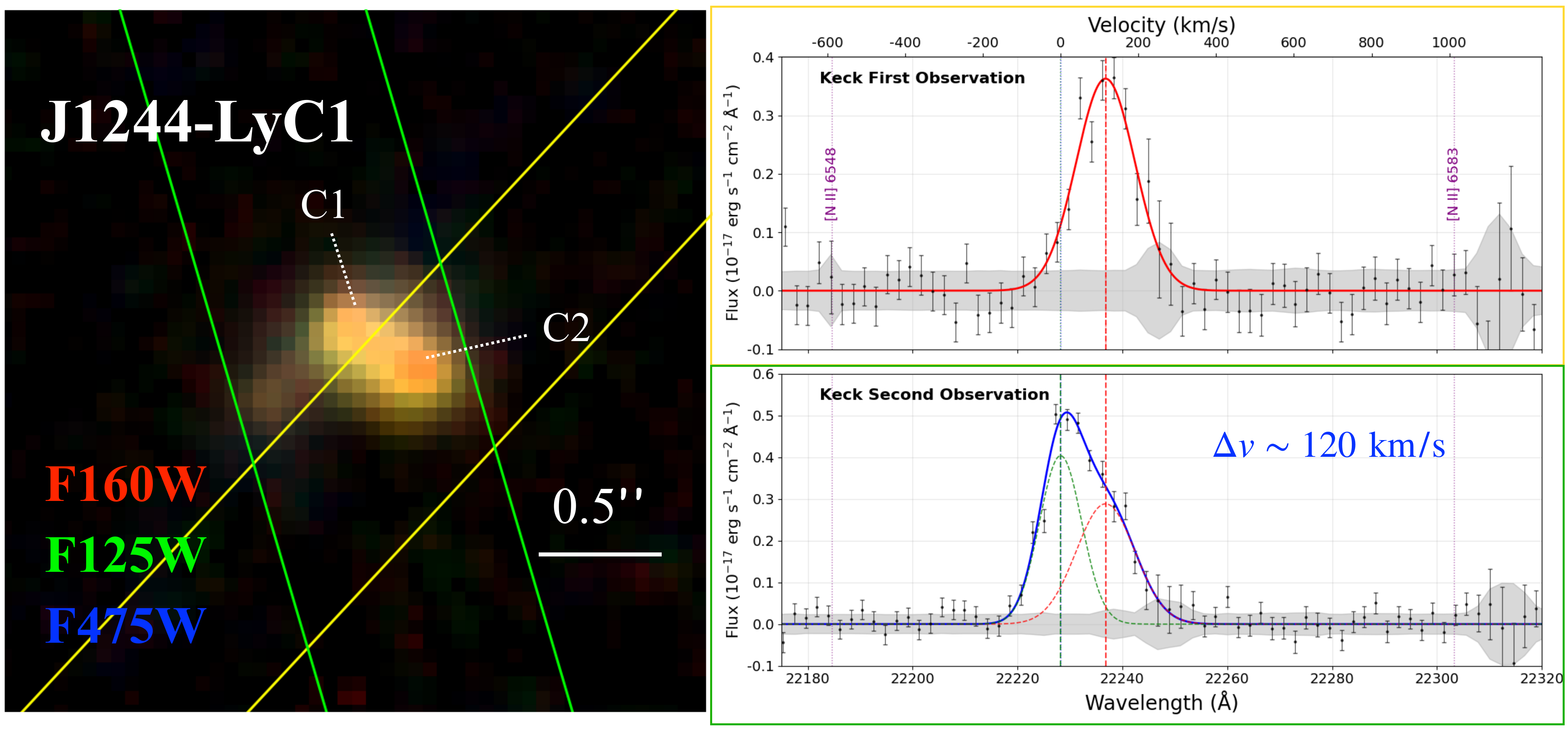}
    % \caption{
    % Keck/MOSFIRE $K$ band spectroscopy (R$\sim3600$) confirming the slightly resolved velocity components of J1244-LyC1 in the H$\alpha$ emission line. 
    % {\sc Left:} Slit positions from two MOSFIRE observations overlaid on the pseudo-color image of J1244-LyC1. 
    % The yellow and green rectangles correspond to the slit orientations for the first and second observations, respectively, and match the colors used for the spectra on the right. 
    % {\sc Right:} H$\alpha$ emission-line profiles and Gaussian fitting results. 
    % For the first observation, a single-Gaussian model is adopted, and the derived line centroid is used to constrain the double-Gaussian fit in the second observation. 
    % Because the blue component is closer to the systemic redshift of J1244-LyC1 ($z=2.387$), we adopt its centroid as the velocity zero-point to illustrate the relative velocity offset between the two components. 
    % The expected locations of the \NII\ $\lambda\lambda6548,6583$ lines are marked; no significant \NII\ emission is detected in either observation.
    % } \label{fig:LyC_2}
    \caption{Keck/MOSFIRE $K$ band spectroscopy ($R\sim3600$) revealing two slightly resolved kinematic components in the H$\alpha$ emission line of J1244-LyC1. {\sc Left:} Slit positions from two MOSFIRE observations overlaid on the pseudo-color image of J1244-LyC1. The yellow and green rectangles correspond to the slit orientations for the first and second observations, respectively, and match the colors used for the spectra on the right. {\sc Right:} H$\alpha$ emission-line profiles and Gaussian fitting results. For the first observation, a single-Gaussian model is adopted, and its centroid and FWHM are used to constrain one component of the double-Gaussian fit in the second observation. This constrained fit yields $\Delta v=116\pm3.6~\mathrm{km\,s^{-1}}$, where the quoted uncertainty is the formal statistical uncertainty conditional on the adopted constraints. Based on the additional fitting tests described in Section~\ref{subsubsect:Ha_D}, we adopt a characteristic velocity separation of $\Delta v\simeq(120\pm10)~\mathrm{km\,s^{-1}}$ for the interpretation. Because the blue component is closer to the systemic redshift of J1244-LyC1 ($z=2.387$), we adopt its centroid as the velocity zero-point to illustrate the relative velocity offset between the two components. The expected locations of the \NII\ $\lambda\lambda6548,6583$ lines are marked; no significant \NII\ emission is detected in either observation.} \label{fig:LyC_2}
\end{figure*}

% \subsubsection{H$\alpha$ Kinematics} \label{subsubsect: Ha_D }

% The two Keck observations provide a clear separation of the double velocity components in the H$\alpha$ emission.  
% As shown in Fig~\ref{fig:LyC_2}, the first observation primarily samples one of the two photometric centers and is dominated by the redshifted H$\alpha$ component.  
% The second observation exhibits a substantial shift in the emission peak relative to the first spectrum.

% We fitted the H$\alpha$ line in the first observation with a single Gaussian initially, then used its centroid and FWHM as constraints for one of the components in a double-Gaussian model for the second spectrum.
% This yields a relative velocity difference of $116 \pm 3.6\,\mathrm{km\,s^{-1}}$ between the two components.

% Because the first MOSFIRE observation is seeing-limited and only partially covers the system, we cannot reliably quantify the flux contribution from the second component in that exposure.  
% Thus, the adopted H$\alpha$ fluxes for the blue and red components come exclusively from the second observation:
% $F_{\mathrm{H}\alpha,\mathrm{red}} = (4.1 \pm 0.32) \times 10^{-17}~\mathrm{erg\,s^{-1}\,cm^{-2}}$, $F_{\mathrm{H}\alpha,\mathrm{blue}} = (4.1 \pm 0.25) \times 10^{-17}~\mathrm{erg\,s^{-1}\,cm^{-2}}.
% $

\subsubsection{H$\alpha$ Kinematics}
\label{subsubsect:Ha_D}

As shown in Fig.~\ref{fig:LyC_2}, the two Keck/MOSFIRE observations provide evidence for two velocity components in the H$\alpha$ emission. We initially fitted the H$\alpha$ line in the first observation with a single Gaussian and subsequently used its centroid and FWHM to constrain one component of a double-Gaussian model for the second spectrum. Under this constrained parameterization, we obtain a relative velocity separation of $\Delta v = 116 \pm 3.6~\mathrm{km\,s^{-1}},$
where the quoted uncertainty is the formal statistical uncertainty conditional on the adopted constraints.

Because the first MOSFIRE observation is seeing-limited and only partially samples the system, the flux contribution of the second velocity component in this exposure cannot be robustly determined. We therefore adopt the H$\alpha$ fluxes of the blue and red velocity components from the second observation alone. After applying the global slit-loss correction for the second observation, we obtain
$F_{\mathrm{H}\alpha,\mathrm{red}}=(4.1 \pm 0.32)\times10^{-17}~\mathrm{erg\,s^{-1}\,cm^{-2}}$ and $F_{\mathrm{H}\alpha,\mathrm{blue}}=(4.1 \pm 0.25)\times10^{-17}~\mathrm{erg\,s^{-1}\,cm^{-2}}.$
These fluxes refer to the two velocity components in the adopted double-Gaussian decomposition and should not be interpreted as unique measurements of the intrinsic H$\alpha$ fluxes of the two photometric components, because of uncertainty in what their exact line shapes should be.

J1244--LyC1 consists of two compact photometric components, C1 and C2 (see Section~\ref{subsect: GALFIT}). The nominal footprint of the first MOSFIRE slit directly intersects primarily C2. However, because the separation between C1 and C2 is smaller than the ground-based seeing, the first spectrum is not expected to isolate C2 completely. To assess this effect, we convolved the \texttt{GALFIT} model of the F160W morphology with the seeing appropriate for each observation and applied the corresponding slit geometries (see Appendix~\ref{app:slit_test}). The resulting component-wise throughputs indicate that approximately 37\% of the total F160W flux of C1 and 65\% of that of C2 enter the first slit. For the second observation, the corresponding throughputs are approximately 78\% and 86\%, respectively. These percentages describe the fraction of each component's own flux entering the slit; they do not directly specify the relative C1 and C2 contributions to the observed H$\alpha$ spectrum.

In principle, the two observations could be treated as two different linear mixtures of the same intrinsic C1 and C2 emission-line profiles, potentially allowing their individual H$\alpha$ fluxes, velocity centroids, and velocity dispersions to be recovered simultaneously. In practice, however, such a decomposition is highly sensitive to the adopted seeing, source centering, slit placement, and relative flux calibration. More importantly, the F160W continuum morphology may not trace the spatial distribution of the H$\alpha$ emission, and the two-component GALFIT model does not include possible line emission associated with the diffuse or tidal structures. We therefore do not use the continuum-based slit throughputs as exact priors on the H$\alpha$ decomposition.

Instead, we conservatively interpret the spectra in terms of the directly observed kinematics. The broader spatial coverage of the second observation reveals a second velocity component that is not significantly required by the first spectrum. Its appearance is naturally consistent with an enhanced contribution from the part of the system, including C1, that is not directly intersected by the nominal footprint of the first slit. Nevertheless, because of seeing convolution and the possible contribution of tidal emission, we cannot establish a unique one-to-one association between each Gaussian velocity component and C1 or C2.

% xxx state the absolute accuracy of the MOSFIRE wavelength scales, determined by OH lines since that determines how precisely the First and Second spectrum can be matched up XXX
As a robustness test, we also fitted the second spectrum with a double-Gaussian model in which both centroids and widths were allowed to vary freely, without using the first observation as a prior. The free double-Gaussian model is strongly preferred over a single-Gaussian model, with $\Delta\mathrm{BIC}\simeq28$, and gives
$ \Delta v = 124^{+7}_{-10}~\mathrm{km\,s^{-1}}.$
A simultaneous fit to both spectra with shared kinematics but independent component amplitudes gives $\Delta v = 115^{+9}_{-10}~\mathrm{km\,s^{-1}}.$
Both tests are consistent with the original constrained value of approximately $116~\mathrm{km\,s^{-1}}$, demonstrating that the detection and approximate separation of the two velocity components are robust. We therefore adopt a characteristic velocity separation of approximately $120~\mathrm{km\,s^{-1}}$, with an uncertainty of order $10~\mathrm{km\,s^{-1}}$. 
The wavelength calibration was determined using sky OH emission lines, with a calibration uncertainty negligible for the velocity offsets discussed here.

We emphasize, however, that the current data do not permit a unique and robust determination of the intrinsic H$\alpha$ luminosity, line profile, or velocity dispersion associated individually with C1 and C2. Spatially resolved spectroscopy would be required to establish such an association.

\subsubsection{Dust Extinction} \label{subsubsect: Dust}

Given detections of both H$\alpha$ and H$\beta$, we estimated the nebular reddening using the Balmer decrement assuming a Milky Way extinction curve:
\begin{equation}
E(B-V) = \frac{2.5}{k_{\mathrm{H}\beta} - k_{\mathrm{H}\alpha}} 
\log_{10}\left( \frac{(H\alpha / H\beta)}{R_0} \right),
\end{equation}
where $R_0 = 2.86$ for Case~B recombination at $T_{\mathrm{e}} = 10^4$\,K and $n_{\mathrm{e}} = 10^2~\mathrm{cm^{-3}}$ \citep{Osterbrock_2006_ISM}.
Here, $k_{\mathrm{H}\alpha}$ and $k_{\mathrm{H}\beta}$ denote the values of the adopted Milky Way reddening curve evaluated at the wavelengths of H$\alpha$ and H$\beta$, respectively.

This yields an integrated average value of $E(B-V) = 0.15^{+0.27}_{-0.22}$.  
Because H$\beta$ is detected only at $4.3\sigma$, the spectroscopic reddening remains uncertain, and we instead adopt the more precise SED-based estimate of $E(B-V) = 0.21 \pm 0.01$ (Section~\ref{subsect: SED}).

\subsubsection{Metallicity} \label{subsubsect: Metallicity}

Although \OII\ is the only strong metal line detected, we estimate a lower limit on the gas-phase metallicity using the \OII/H$\beta$ relation from \citet{sanders_2025}:
\begin{equation}
\log\left(\frac{[\mathrm{O\,II}]}{H\beta}\right) = 
0.172 + 0.954 \cdot x - 0.832 \cdot x^{2},
\end{equation}
where $x = 12 + \log(\mathrm{O/H}) - 8$, and 
$\log([\mathrm{O\,II}]/H\beta) = 0.43^{+0.12}_{-0.10}$.

The \OII/H$\beta$ relation is calibrated and valid only over 
$12 + \log(\mathrm{O/H}) \in [7.3, 8.6]$, corresponding to 
$\log([\mathrm{O\,II}]/H\beta) \lesssim 0.5$.  
Since our measured ratio exceeds this threshold at the $1\sigma$ upper bound, the relation cannot provide a reliable metallicity upper limit.  
We therefore adopt the $1\sigma$ lower bound of the measured ratio, which corresponds to 
$12 + \log(\mathrm{O/H}) > 8.19$, as a conservative lower limit for constraining the SED-fitting parameter space.
The final SED-fitting results are fully consistent with this metallicity lower limit derived from the emission line fluxes (see Table~\ref{tab:SED_result})

Furthermore, while our previous work has examined the resolved metallicity map in this MAMMOTH field \citep{Li_22} for galaxies with securely detected \OIII\ emission lines, the absence of a robust \OIII\ line map for this target prevents a reliable metallicity map using joint strong-line constraints (i.e., \OIII and \OII). Due to the low SNR for each \OII and H$\beta$ spaxel, we did not manage to calculate a metallicity map from them.
\subsubsection{Star Formation} \label{subsubsect: SF}

With the adopted extinction correction, the dust-corrected H$\alpha$ luminosity implies $\mathrm{SFR(H\alpha)} \approx 48 \pm 2.3~M_\odot\,\mathrm{yr^{-1}}$ using the \citet{Kennicutt_1998} calibration.  
This value agrees with the SED-derived SFR (Table~\ref{tab:SED_result}).  
Given its stellar mass of $\sim 10^{10.2}~M_\odot$, the SFR is slightly above that of typical main-sequence galaxies at comparable redshift \citep{Speagle_2014_MS}. %consistent with a merger-driven starburst phase.

% The WFC3/G141 grism data provide a spatially resolved \OII\ map (Fig.~\ref{fig:5}). 
% Although the absence of a resolved metallicity map prevents a quantitative conversion to a spatially resolved SFR map \citep{Kewley_2002}, the \OII\ distribution still traces recent star formation on $\sim$3–10\,Myr scales \citep{Kennicutt_2012_SFR}.  
% We examine its connection to LyC escape in Section~\ref{subsect: S_fesc_c}.

The spatial distribution of the SFR on different timescales in J1244-LyC1 is important for our analysis. The WFC3/G141 grism data provide spatially resolved maps of both \OII (Fig.~\ref{fig:5}) and H$\beta$. However, the H$\beta$ map has insufficient SNR, preventing a reliable SFR map derived directly from Balmer emission. Constructing an SFR map from the \OII\ emission would require a resolved metallicity map \citep{Kewley_2002}, but as discussed in Section~\ref{subsubsect: Metallicity}, the current data do not support a robust spatially resolved metallicity estimate.

Although we cannot derive a full SFR map, the spatial distribution of \OII\ nevertheless traces recent star formation on $\sim$3--10\,Myr timescales \citep{Kennicutt_2012_SFR}. We examine its connection to LyC escape in Section~\ref{subsect: S_fesc_c}.

\subsection{SED Model Fits} \label{subsect: SED}

\begin{figure*}[htbp]
    \centering
    \includegraphics[width=1\textwidth]{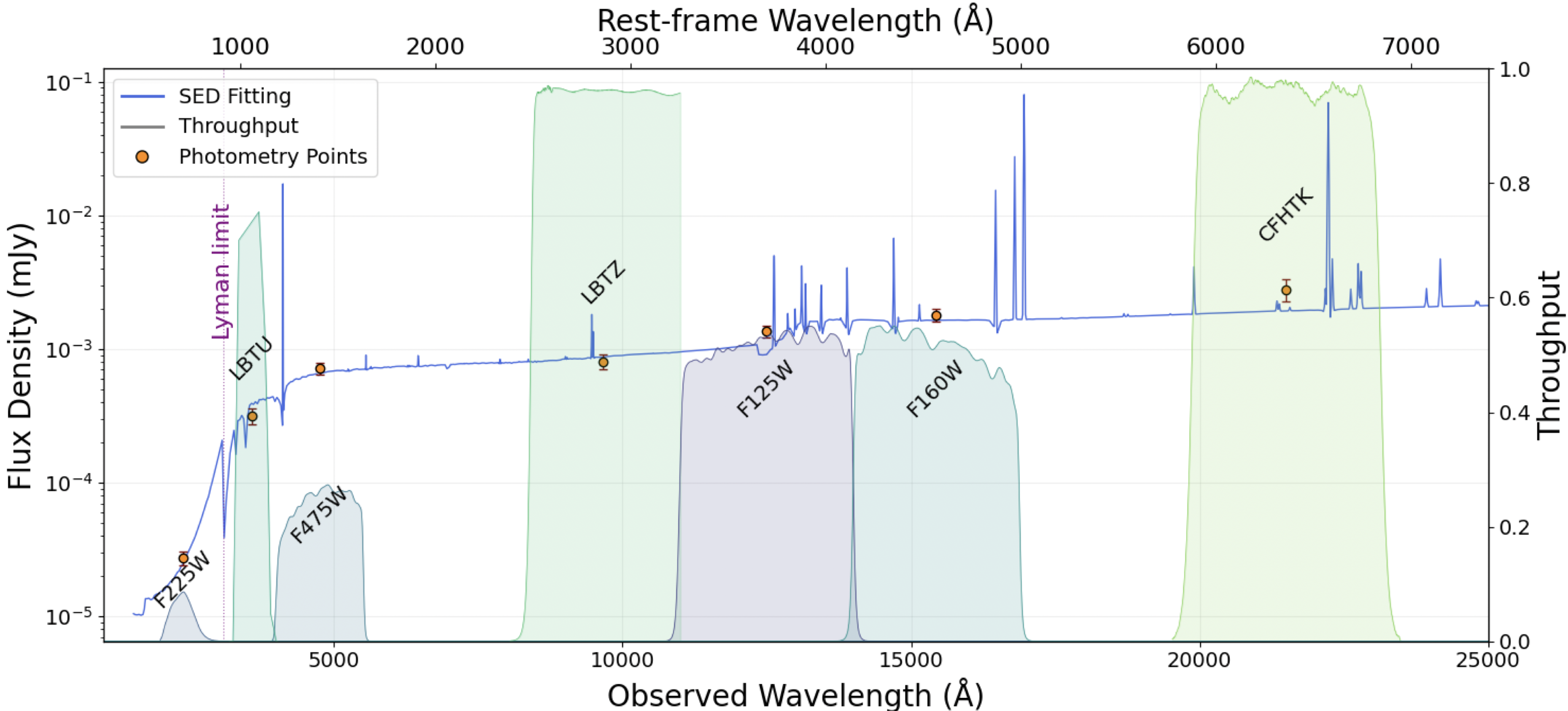}
    \caption{
    % The best-fit SED model (blue) of J1244-LyC1 at $z=2.39$ using \texttt{CIGALE}, fit to the existing broad-band photometry covering a wide wavelength range of $600$--$7000\,\AA$ in the rest-frame.
    % The SED fitting results are shown in Table~\ref{tab:SED_result}, and the multi-band photometry is shown in Table~ \ref{tab:measure}.
    % The SED of J1244-LyC1 is dominated by a young stellar population with an age  $330 \pm 120$ Myr and a recent SFR of $65 \pm 33\,M_\odot\,\mathrm{yr^{-1}}$.
    The best-fit SED model (blue) of J1244-LyC1 at $z=2.39$, obtained with \texttt{CIGALE} using the available broad-band photometry covering a rest-frame wavelength range of $600$--$7000\,$\AA. The photometric measurements are listed in Table~\ref{tab:measure}. The physical parameters inferred from the marginalized CIGALE model-grid distributions, together with their corresponding $1\sigma$ uncertainties, are summarized in Table~\ref{tab:SED_result}. The SED is dominated by a young stellar population with an age of $330\pm120$ Myr and a recent SFR of $65\pm33\,M_\odot\,\mathrm{yr}^{-1}$.
    % \vspace{-.5em}
    }\label{fig:3} 
\end{figure*}

We performed SED fitting to constrain the physical properties of J1244-LyC1, which provides the necessary stellar population information for estimating $f_{\mathrm{esc}}$. The fitting was carried out using the latest version of {\tt CIGALE} (v2025; \citealt{Boquien_cigale_2019}), and the resulting best-fit SED is shown in Fig.~\ref{fig:3}. We adopted the \citet{Bruzual_2003_BC03} stellar population synthesis (BC03) model, a delayed exponentially declining star formation history, and the \citet{Calzetti_2000} dust attenuation law. The escape fraction $f_{\mathrm{esc}}$ was included as a free parameter. 

The absorption in the $U$ band caused by the blended Lyman-series lines is $\sim 12\%$, as estimated using the IGM  model adopted in {\tt CIGALE} \citep{Meiksin_2006}.

% The absorption in the u-band from the blended Lyman series lines was 

The photometric measurements used in the SED fitting include the LBT $U$ and $Z$ bands, the CFHT $Ks$ band, and the HST/WFC3 F225W (LyC), F475W, F125W, and F160W bands. All HST images were PSF-matched to the F160W image to ensure consistent photometry. For the ground-based data, we used the F160W image as a high-resolution prior and employed \textsc{TPHOT} to obtain accurate flux measurements \citep{Yiming_2025}. The results are summarized in Table~\ref{tab:measure}.

% As discussed in Section~\ref{subsect: Spec_A}, we derived a lower limit on metallicity as well as estimates of $E(B-V)$ and $\mathrm{SFR}_{\mathrm{H\alpha}}$. These spectroscopic measurements were incorporated as constraints in the SED fitting. xxx really? i dont see how the Balmer decrement very rough guestimate of EB-V~0.15 is relevant to this SED fit. CIGALE actually wanted a somewhat LARGER stellar reddening than gas reddening which is not generally seen. but the Balmer uncertainties are so large that anything goes, the gas EB-V seems irrelevant XXX The derived physical parameters are listed in Table~\ref{tab:SED_result}.
As discussed in Section~\ref{subsect: Spec_A}, we derived a lower limit on the metallicity, as well as estimates of the nebular reddening and $\mathrm{SFR}_{\mathrm{H\alpha}}$. The metallicity and H$\alpha$-based SFR measurements were used to inform the SED fitting, while the nebular reddening estimate was not imposed as a constraint because of its large uncertainty. The derived physical parameters are listed in Table~\ref{tab:SED_result}.

% \begin{deluxetable*}{lcccccccc}
% \tablecaption{Physical properties of J1244-LyC1 derived from SED fitting \label{tab:SED_result}}
% \tablehead{
% \colhead{RA} &
% \colhead{DEC} &
% \colhead{$z_{\mathrm{spec}}$} &
% \colhead{$\log(M_{\ast}/M_{\odot})$} &
% \colhead{SFR$_{\mathrm{H}\alpha}$ (M$_{\odot}$\,yr$^{-1}$)} &
% \colhead{SFR$_{\mathrm{SED}}$ (M$_{\odot}$\,yr$^{-1}$)} &
% \colhead{$E(B-V)_{\mathrm{gas}}$} &
% \colhead{$12+\log(\mathrm{O/H})$} &
% \colhead{$f_{\mathrm{esc}}$}
% }
% \startdata
% 190.87348 & 35.90626 & 2.387 &
% $10.15^{+0.19}_{-0.35}$ &
% $48\pm2$ &
% $65\pm33$ &
% $0.21\pm0.01$ &
% $8.47^{+0.11}_{-0.14}$ &
% $0.37\pm0.08$ \\
% \enddata
% \tablecomments{
% All reported uncertainties represent 1$\sigma$ errors.
% }
% \end{deluxetable*}
\begin{deluxetable*}{lcccccccc}
\tablecaption{Physical properties of J1244-LyC1 derived from the SED fitting and spectroscopic analysis}  \label{tab:SED_result}
\tablehead{
\colhead{RA} &
\colhead{DEC} &
\colhead{$z_{\mathrm{spec}}$} &
\colhead{$\log(M_{\ast}/M_{\odot})$} &
\colhead{SFR$_{\mathrm{H}\alpha}$ ($M_{\odot}$\,yr$^{-1}$)} &
\colhead{SFR$_{\mathrm{SED}}$ ($M_{\odot}$\,yr$^{-1}$)} &
% \colhead{Age (Myr)} &
\colhead{$E(B-V)_{\mathrm{gas}}$} &
\colhead{$12+\log(\mathrm{O/H})$} &
\colhead{$f_{\mathrm{esc}}$}
}
\startdata
190.87348 & 35.90626 & 2.387 &
$10.15^{+0.19}_{-0.35}$ &
$48\pm2$ &
$65\pm33$ &
% $330\pm120$ &
$0.21\pm0.01$ &
$8.47^{+0.11}_{-0.14}$ &
$0.37\pm0.08$ \\
\enddata
\tablecomments{
The quoted uncertainties for the \texttt{CIGALE}-derived parameters are obtained from their marginalized likelihood distributions over the full model grid and represent the corresponding $1\sigma$ intervals. The quantities SFR$_{\mathrm{H}\alpha}$ is derived independently from the spectroscopic analysis and were incorporated as external constraints in the SED fitting.
}
\end{deluxetable*}

\subsection{Decomposing Images Using \texttt{GALFIT}} \label{subsect: GALFIT}

Across all available imaging bands, J1244-LyC1 exhibits two bright components and tidal features (see Fig.~\ref{fig:1}). 
To characterize its structural properties, we performed two-dimensional surface brightness modeling using {\tt GALFIT}.

\begin{figure*}[htbp] 
    \centering
    \includegraphics[width=1\textwidth]{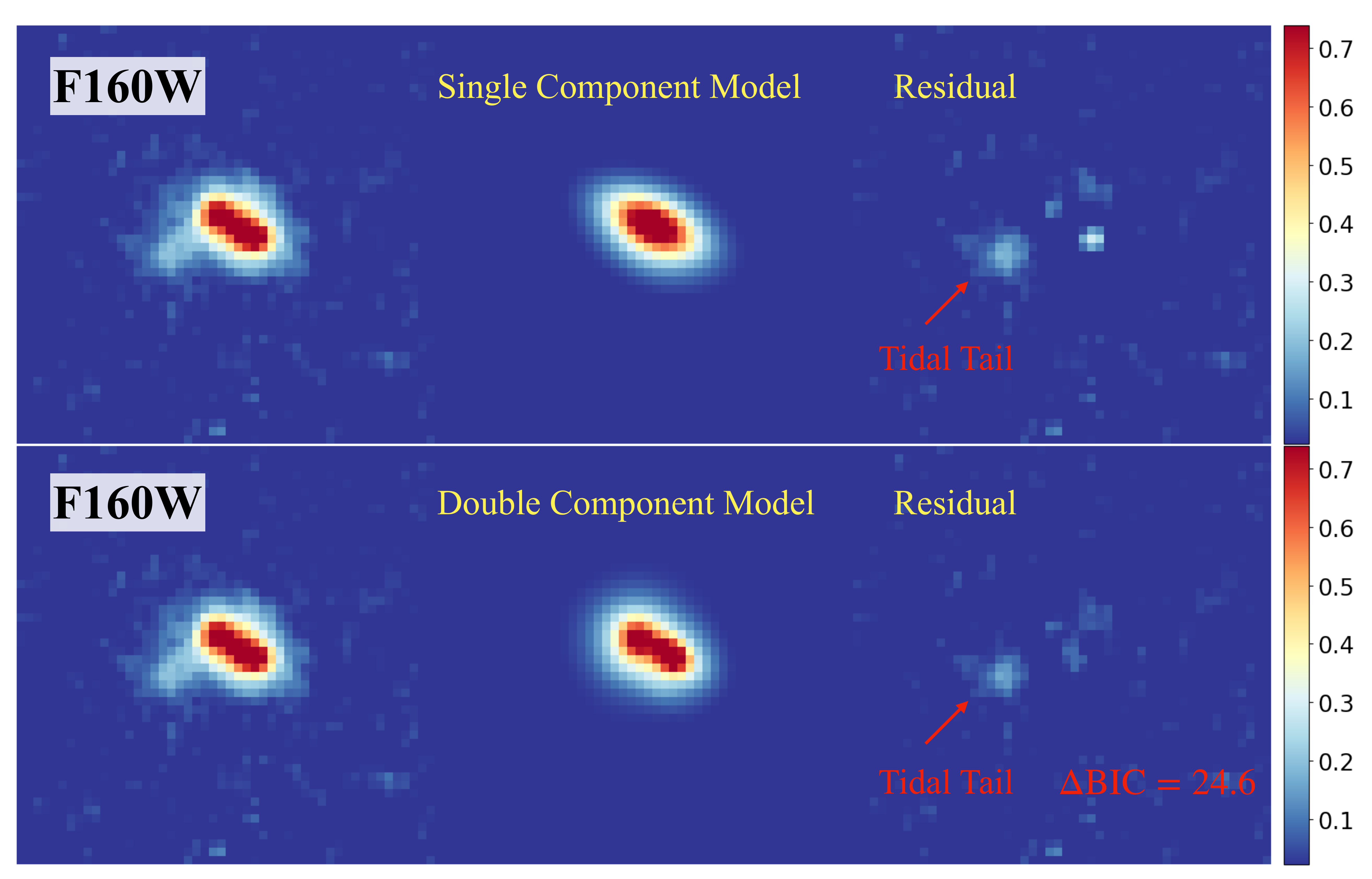}
    \caption{
    {\tt GALFIT} decomposition of J1244-LyC1 in HST F160W imaging. 
    All images are normalized to the peak value of the observed image. 
    Single-component ({\sc top}) and Double-component ({\sc bottom}) {\tt GALFIT} models for the F160W image. 
    The Double-component fit provides a significantly better description of the main body of J1244-LyC1, whereas the single-component model leaves behind prominent structural residuals. 
    The red arrows in the residual maps highlight an extended residual feature, which may trace tidally disturbed structure associated with the merger.
    % The red arrows in the residual map highlight the tidal-tail structure, a signature of galaxy mergers.  xxx but it looks more like a ROUND BLOB, unlike a 'tail' which would be quite ELONGATED along its orbit path XXX
    % {\sc Bottom:} Double-component GALFIT decomposition for the F475W image. 
    % Owing to the higher spatial resolution of F475W relative to F160W, the residual map reveals clear tidal-tail features, further supporting the merger nature of the system.
    % \vspace{-0.5em}
    \label{fig:GALFIT}
    } 
\end{figure*}

As illustrated in Fig.~\ref{fig:GALFIT}, we modeled the F160W image, which provides the highest spatial resolution at a wavelength closest to the Keck/MOSFIRE $K$ band used for the H$\alpha$ observations. This allows a direct comparison between the morphological substructures and the two kinematic components revealed in the H$\alpha$ emission.

We carried out both single-Sérsic and double-Sérsic fits. The single-component model leaves substantial residuals, whereas the double-component model provides a significantly improved description of the global morphology. The total $\chi^2$ decreases from 1706.0 to 1642.3, corresponding to reduced $\chi^2$ values of 0.684 and 0.660, respectively. After accounting for the additional free parameters, the double-component model is still strongly preferred, with $\Delta\mathrm{BIC}=24.6$. The residual maps further highlight extended tidal structures.

As an independent non-parametric assessment, we measured the morphology of J1244-LyC1 using segmentation map, obtaining a Gini coefficient of $G=0.61$ and $M_{20}=-1.46$ \citep{Lotz_2004}. These values place J1244-LyC1 in the merger region of the $G$--$M_{20}$ diagram \citep{Lotz_2008}, further supporting its disturbed, multi-component morphology.

% The two components, labeled C1 and C2, have a flux ratio of roughly 1.8:1 and are separated by 2.5\,kpc (0.3\arcsec). Only one component falls within the slit during the first Keck pointing.
% % Combined with the dual H$\alpha$ velocity components (Section~\ref{subsect: Spec_A}), which exhibit a relative velocity offset of $116 \pm 3.6~\mathrm{km\,s^{-1}}$, these results confirm that J1244-LyC1 is a late-stage major merger.
% % xxx explain how you know what STAGE it is--is this before or after the FIRST closest approach, or after the second and final merger. there is no way to tell without some detailed model. and what if there is a LARGE TRANSVERSE VELOCITY DIFFERENCE we dont know which might be a lot more than 120km/sec? XXX
% Combined with the dual H$\alpha$ velocity components (Section~\ref{subsect: Spec_A}), which exhibit a line-of-sight velocity offset of $116 \pm 3.6~\mathrm{km,s^{-1}}$, the small projected separation and disturbed tidal morphology strongly favor a morphologically late-stage major-merger interpretation for J1244-LyC1. The precise merger stage, however, cannot be uniquely determined because the transverse velocity is unconstrained and no detailed dynamical model is available.

The two components, labeled C1 and C2, have an F160W flux ratio of roughly 1.8:1 and are separated by a projected distance of 2.5\,kpc ($0.3\arcsec$). The nominal footprint of the first Keck slit primarily intersects C2, although both components contribute after seeing convolution (Section~\ref{subsubsect:Ha_D}).
% Appendix~\ref{app:slit_test}).

Together with the two H$\alpha$ kinematic components, which have an adopted characteristic line-of-sight velocity separation of approximately $120~\mathrm{km\,s^{-1}}$ with an uncertainty of order $10~\mathrm{km\,s^{-1}}$, the small projected separation and disturbed morphology support a morphologically late-stage major-merger interpretation for J1244-LyC1. 
The precise merger stage, however, cannot be uniquely determined because the transverse velocity is unconstrained and no detailed dynamical model is available.

\subsection{LyC Escape Fraction} \label{subsect: fesc}

The escape fraction $f_{\mathrm{esc}}$ is constrained observationally by comparing the observed LyC flux to the intrinsic LyC flux expected from stellar population models. Several methods have been developed \citep{Jaskot_lowz_review_2025}:  
(1) using the H$\beta$ line to quantify nebular absorption of LyC photons (e.g., \citealt{Izotov_2016_nature,Flury_2022_I});  
(2) fitting the multi-band SED while allowing $f_{\mathrm{esc}}$ to vary as a free parameter (e.g., \citealt{izotov_detection_2016,Fletcher_2019}); and  
(3) determining the relative escape fraction based on the ratio of ionizing to non-ionizing fluxes, which can be corrected for dust attenuation to obtain the absolute escape fraction (e.g., \citealt{Steidel_2001,wang_lyman_2025,gupta_mosel_2024,steidel_keck_2018}).

Although J1244-LyC1 shows an H$\beta$ detection in the HST/WFC3 G141 grism spectrum, the H$\beta$–$f_{\mathrm{esc}}$ method is not applicable because of the following reasons. 
First, estimating the nebular absorption of LyC photons using H$\beta$ still relies on the assumptions of isotropy and a uniform stellar population \citep{Izotov_2016_nature, Flury_2022_I}. 
Given the pronounced spatial inhomogeneity of the LyC signal in J1244-LyC1  (see Section~\ref{subsect: S_fesc_c}), we expect that the galaxy-integrated $f_{\mathrm{esc}}$ inferred from H$\beta$ would significantly deviate from the true value. 
Second, J1244-LyC1 is substantially more dusty than most LyC candidate samples, and H$\beta$ only constrains the nebular absorption of LyC photons, providing no information on the additional attenuation by dust\citep{Jaskot_lowz_review_2025}. 
This makes the method particularly unsuitable for J1244-LyC1. 
Finally, the SNR of H$\beta$ is relatively low, which prevents us from placing strong constraints on $f_{\mathrm{esc}}$ using this approach.

% The SED-based analysis (Section~\ref{subsect: SED}) yields $f_{\mathrm{esc}}\simeq37\%$.

We adopt the third approach to estimate the escape fraction. The relative escape fraction is commonly defined between the rest-frame LyC and UV bands as
\begin{equation}
    f_{\mathrm{esc,rel}} = 
    \frac{(L_{\mathrm{UV}}/L_{\mathrm{LyC}})_{\mathrm{int}}}
         {(F_{\mathrm{UV}}/F_{\mathrm{LyC}})_{\mathrm{obs}}}
    \times t_{\mathrm{IGM}}^{-1},
\end{equation}
where $(L_{\mathrm{UV}}/L_{\mathrm{LyC}})_{\mathrm{int}}$ is the intrinsic UV-to-LyC luminosity ratio (typically evaluated at $\lambda_{\mathrm{LyC}}=900$\,\AA\ and $\lambda_{\mathrm{UV}}=1500$\,\AA),  $t_{\mathrm{IGM}}=e^{-\tau_{\mathrm{IGM}}}$ is the transmitted fraction of ionizing photons through the IGM, and  $\tau_{\mathrm{IGM}}$ is the IGM opacity (e.g., \citealt{Steidel_2001, inoue_updated_2014, Meiksin_2006}).

For J1244-LyC1, we compute the relative escape fraction using the observed photometry in the HST/WFC3 UVIS F225W and F475W filters. We treat the F225W band (central wavelength $\sim 665$\,\AA; $F_{\mathrm{F225W}} = 27.81^{+0.20}_{-0.17}$~AB mag) as the LyC band, and the F475W band (central wavelength $\sim 1400$\,\AA; $F_{\mathrm{F475W}} = 24.27^{+0.01}_{-0.01}$~AB mag) as the non-ionizing UV band. 
The intrinsic luminosity ratio, $(L_{\mathrm{F475W}}/L_{\mathrm{F225W}})_{\mathrm{int}} = 7.5$, is obtained directly from our best-fit stellar population models.

To assess the robustness of this ratio, we repeated the SED fitting using 1000 Monte Carlo realizations of the photometry after excluding the F225W measurement. The resulting distribution gives $(L_{\mathrm{F475W}}/L_{\mathrm{F225W}})_{\mathrm{int}}=7.71^{+2.04}_{-0.38}$, fully consistent with the value derived from the original fit and indicating that the intrinsic ratio is not driven by the observed LyC-band flux. We also tested reasonable alternative SFHs and found that the inferred ratio remains consistent within the uncertainties. A larger systematic difference arises when adopting the BPASS binary stellar population models \citep{Stanway_2018}, which yield an intrinsic ratio approximately 26\% lower than that obtained with the fiducial BC03 single-star models. This difference would reduce the inferred $f_{\mathrm{esc}}$ by the same proportion. Since the available observations do not provide sufficient leverage to distinguish between the two stellar population models, we retain the BC03 single-star result as our fiducial value while noting the systematic uncertainty associated with binary stellar evolution.

The filter throughput weighted IGM transmission, $\bar{t}_{\mathrm{IGM}} $ is computed using the IGM absorption prescription of \citet{Meiksin_2006}, which is also adopted by {\tt CIGALE}. Following the method of \citet{wang_lyman_2025}, we calculate the transmission coefficient for each wavelength and then perform a transmission-weighted average over the F225W bandpass:
% \begin{equation}
%     \langle t_{\mathrm{IGM}}\rangle_{\mathrm{F225W}} = 
%     \frac{\int T_{\mathrm{F225W}}(\lambda)\, e^{-\tau_{\mathrm{IGM}}(\lambda)}\, d\lambda}
%     {\int T_{\mathrm{F225W}}(\lambda)\, d\lambda},
% \end{equation}
\begin{equation}
\bar{t}_{\mathrm{IGM}} =
\frac{
\int e^{-\tau_{\mathrm{IGM}}}\, \frac{T_{\mathrm{F225W}}}{\lambda}\, d\lambda
}{
\int \frac{T_{\mathrm{F225W}}}{\lambda}\, d\lambda
}
\end{equation}
which yields $\bar{t}_{\mathrm{IGM}}\approx 0.12$.

After applying the dust correction (as detailed in Section~\ref{subsubsect: Dust}, adopting $E(B-V)=0.21$), the absolute escape fraction is derived as
\begin{equation}
    f_{\mathrm{esc,abs}} = f_{\mathrm{esc,rel}} \times 10^{-0.4\,A_{UV}}.
\end{equation}
By following the Calzetti dust attenuation law \citep{Calzetti_2000} appropriate for high-z star-forming galaxies, we adopted $A_{UV} = 10.33\times E(B-V)$ and calculated $f_{\mathrm{esc,abs}} = 0.36\pm 0.06$, which is consistent with the SED result (see Table~\ref{tab:SED_result}).
In the limiting case of no IGM attenuation ($\bar{t}_{\mathrm{IGM}}=1$), the inferred escape fraction decreases to $f_{\mathrm{esc,abs}}=0.043\pm0.007$, providing a strict lower bound under the most transparent possible line of sight.
% where $A_{UV} = 10.33\,E(B-V)$ following the Calzetti dust attenuation law \citep{Calzetti_2000} appropriate for high-z star-forming galaxies.
% We calculate $f_{\mathrm{esc,abs}} = 0.36\pm 0.03$, which is consistent with the SED-based estimate. 

\section{DISCUSSION} \label{sect: discussion}
In this section, we examine the uncertainties of $f_{\mathrm{esc}}$ calculation, analyze the spatially resolved LyC escape to understand the nature of the massive LyC leakers, and discuss the environmental effects on the LyC leakage.

\subsection{$f_{\mathrm{esc}}$ Calculation} \label{subsect: fesc_c}

For an individual high-redshift galaxy, the estimation of $f_{\mathrm{esc}}$ is inevitably affected by substantial uncertainties introduced by IGM transmission. In practice, we can only approximate this effect using the average IGM transmission predicted by models at the corresponding redshift. However, given the large sightline-to-sightline variations in IGM properties, $f_{\mathrm{esc}}$ and $t_{\mathrm{IGM}}$ are intrinsically degenerate \citep{Jaskot_lowz_review_2025,wang_lyman_2025}. Consequently, $f_{\mathrm{esc}}$ is often overestimated, as LyC leakers are more readily detected along relatively transparent sightlines \citep{Bassett_2021}. This represents one of the most significant challenges in LyC studies at high redshift: inferred values of $f_{\mathrm{esc}}$ always rely on assumed average IGM models \citep{inoue_updated_2014,Meiksin_2006}.

On the other hand, we could directly estimate the $f_{\mathrm{esc}}$ by determining the intrinsic LyC flux, $L_{\mathrm{LyC,int}}$.
% Furthermore, direct estimation of $f_{\mathrm{esc}}$ requires determining the intrinsic LyC flux, $L_{\mathrm{LyC,int}}$.
In principle, if the SED fit is sufficiently robust, $f_{\mathrm{esc}}$ can be computed simply as $L_{\mathrm{LyC,obs}} / L_{\mathrm{LyC,int}}$. However, as defined in Section~\ref{subsect: fesc}, a commonly adopted approach expresses $f_{\mathrm{esc}}$ in terms of the UV flux, introducing $f_{\mathrm{UV,obs}}$. Because $f_{\mathrm{UV,obs}}$ serves as a proxy for the SFR, this method helps mitigate systematic uncertainties associated with the model dependence of SED fitting \citep{Siana_2007,steidel_keck_2018}.

This definition requires an estimate of the intrinsic stellar ratio $(L_{\mathrm{UV}}/L_{\mathrm{LyC}})_{\mathrm{int}}$, which depends on both the stellar population model inferred from the SED and the specific wavelength ranges used for $L_{\mathrm{LyC,int}}$ and $L_{\mathrm{UV,int}}$. These fluxes are typically defined at 900\,\AA\ and 1500\,\AA\ \citep{Siana_2007,Steidel_2001}. 
% Although \citet{Simmonds_2024} noted that this convention may lead to a systematic overestimate of $f_{\mathrm{esc}}$, it remains practical at high redshift: due to strong IGM attenuation, LyC signals are intrinsically brighter and more detectable near the Lyman limit \citep{inoue_updated_2014}. 
Although \citet{Simmonds_2024} argued that defining the intrinsic ratio using 900\,\AA\ and 1500\,\AA\ can systematically overestimate $f_{\mathrm{esc}}$—and suggested instead adopting fluxes at 700\,\AA\ and 1100\,\AA —this alternative is less suitable at high redshift. Owing to strong IGM absorption, LyC photons near the Lyman limit (e.g., 900\,\AA) remain substantially more detectable than those at shorter wavelengths such as 700\,\AA\ \citep{inoue_updated_2014}. 
Furthermore, the spectral range between the Lyman limit and Ly$\alpha$ is also significantly affected by the IGM; thus, while the traditional definition may overestimate $f_{\mathrm{esc}}$, it does not introduce additional IGM-related uncertainties and is particularly effective for identifying LyC leakers. 
Since high-redshift LyC searches are predominantly imaging-based, the definition also depends on filter choice, as discussed in Section~\ref{subsect: fesc}.

Although the ratio $(L_{\mathrm{UV}}/L_{\mathrm{LyC}})_{\mathrm{int}}$ primarily reflects the underlying stellar population, it can, in principle, span a broad range. Empirically, typical values fall between $3$ and $7$ \citep[e.g.,][]{Guaita_2017,Alavi_2020,Rutkowski_2017,Smith_2018,wang_lyman_2025}. We note that the value derived for J1244-LyC1 slightly exceeds this range, largely because the rest-frame wavelength probed by the F225W filter is bluer and therefore intrinsically fainter than $F_{\mathrm{rest,900}}$.

\subsection{Spatially resolved LyC escape} \label{subsect: S_fesc_c}

\begin{figure*}[htbp]
    \centering
    \includegraphics[width=1\textwidth]{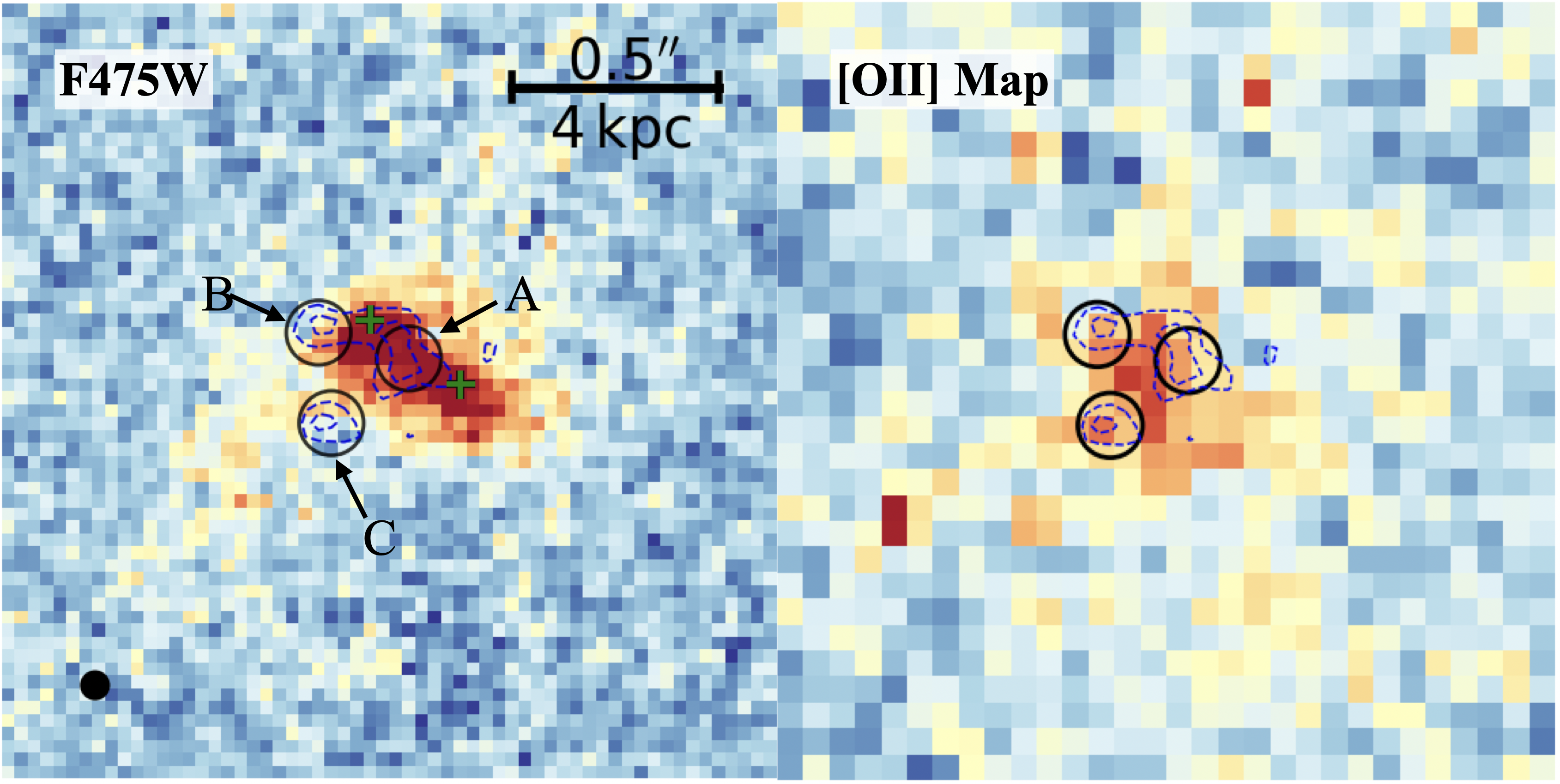}
    \caption{
    UV morphology and \OII\ emission distribution of J1244-LyC1. 
    We present the HST/WFC3 F475W image (rest-frame UV continuum) together with the \OII\ emission map extracted from the HST WFC3/G141 slitless spectroscopic data. 
    The blue dashed contours trace the spatial distribution of the LyC emission. 
    The green crosses mark the two luminosity centers identified in the F475W image. 
    Three black circles (diameter 0.15\arcsec) indicate the locations of the three LyC-emitting clumps defined in our analysis. 
    At all three clump positions, the flux distribution in F475W differs significantly from that in the \OII\ map, highlighting spatial variations in recent ($\sim$3–10 Myr) versus longer-timescale ($\sim$10–100 Myr) star-formation activity.
    % \vspace{-0.5em}
    }
    \label{fig:5}
% \end{figure}
\end{figure*}

The LyC emission of J1244-LyC1 is detected at a significance level of $6\sigma$, and its spatial distribution exhibits a clear clumpy morphology (see Fig.~\ref{fig:5}). This suggests that different LyC-emitting clumps may have distinct formation mechanisms. To quantitatively identify the locations and centroids of these clumps, we employed a growth-curve algorithm. Specifically, we centered circular apertures on different spatial positions and gradually increased the aperture radius to identify independent LyC-emitting regions with ${\rm SNR} > 3$ that do not spatially overlap. Through testing, we found that adopting a maximum radius of 2.5~pixels effectively separates the LyC emission into three independent components—``clumps" A, B, and C—all with ${\rm SNR} > 3$.
The F225W uncertainties used for the clump photometry and SNR selection were corrected using the same empirical procedure, evaluated at the radius of each clump aperture (Section~\ref{subsect: LyC_D}).

In principle, $f_{\mathrm{esc}}$ can be independently calculated for each of the three LyC-emitting clumps, which requires determining their respective intrinsic flux ratios $(f_{\mathrm{UV,int}} / f_{\mathrm{LyC,int}})$ from SED fitting \citep{Siana_2007, steidel_keck_2018}. However, since only four high-resolution imaging bands are available, it is difficult to obtain reliable SED fits at each position, and therefore the derived $f_{\mathrm{esc}}$ values would be highly uncertain.
% To discuss the relative differences among the three clumps, 
We therefore assume that their stellar populations are consistent with that of the global merger system as inferred from the integrated SED, and that the dust distribution is spatially uniform. 
% Under this assumption, we estimate $f_{\mathrm{esc}}$ values of approximately $98$\%, $205$\%, and $330$\% for clumps A, B, and C, respectively.
Under this assumption, we estimate that the $f_{\mathrm{esc}}$ of clump A is around 1, while the $f_{\mathrm{esc}}$ values of clumps B and C are significantly above 1.

It is evident that the $f_{\mathrm{esc}}$ values of clumps A, B, and C are substantially overestimated. 
Overall, the most likely cause of this overestimate is the systematic uncertainty introduced by the choice of IGM model. 
To ensure consistency between the {\tt CIGALE} SED-fitting results and our $f_{\mathrm{esc}}$ calculations, 
we adopted the same IGM model implemented in {\tt CIGALE} \citep{Meiksin_2006}. 
However, as shown in \citet{inoue_updated_2014}, different IGM models exhibit very large discrepancies. 
For J1244-LyC1 at $z=2.387$, the predicted $\bar{t}_{\mathrm{IGM}}$ values vary widely: 
\citet{Meiksin_2006} gives $\sim 0.12$, \citet{steidel_keck_2018} gives $\sim0.26$, and \citet{inoue_updated_2014} gives $\sim0.33$. 
As a result, the inferred $f_{\mathrm{esc}}$ may differ by a factor of $2$--$3$ depending on the adopted IGM model. 
This is one of the primary reasons why the $f_{\mathrm{esc}}$ values of all three clumps are collectively overestimated.

In addition, the three clumps show strong internal differences in their inferred escape fractions: 
clumps B and C exhibit significantly higher $f_{\mathrm{esc}}$ than clump A.
For clump~A, which is spatially coincident with the main stellar body of the galaxy, the assumption of stellar population consistency is most reasonable. Its $f_{\mathrm{esc}}$ is significantly higher than the system-wide average, indicating that the global $f_{\mathrm{esc}}$ likely represents a luminosity-weighted average of regions with locally higher escape fractions. This scenario is consistent with observations and models in which ionizing photons escape through low-$N_{\rm HI}$ sightlines (i.e., holes or channels), while the non-ionizing UV continuum is dominated by the integrated stellar population along the entire line of sight \citep{Jaskot_lowz_review_2025,giovinazzo2025breakingcosmicfogjwstnirspec}.

For clumps~B and~C, our calculations yield results with $f_{\mathrm{esc,abs}}>1$, which are significantly higher than that of clump A. Both regions are spatially offset from the main stellar body and contribute only minor flux fractions in multi-band imaging. Thus, assuming identical stellar populations as the integrated SED fit is likely invalid; their stellar populations are probably much younger. If we instead assume $f_{\mathrm{esc}}=1$ for these two regions, we can constrain their intrinsic stellar flux ratios $(f_{\mathrm{UV}}/f_{\mathrm{LyC}})_{\mathrm{int}}$ to be $\lesssim3.5$ and $\lesssim2.5$, respectively. This implies that the star formation histories in these regions are more bursty and short-lived compared to the overall system. Another important factor is the distribution of dust \citep{Ji_2025}. During the merging process, dust can rapidly enrich and remain spatially inhomogeneous on short timescales, enhancing anisotropic escape of LyC photons \citep{Ejdetj_rn_2025}. Both effects directly impact our estimation of local $f_{\mathrm{esc}}$ values for individual LyC-emitting clumps.

Complementary evidence comes from the \OII\ map. While \OII\ emission traces star formation on short ($\sim$3--10\,Myr) timescales, the UV continuum traces star formation over longer ($\sim$10--100\,Myr) timescales \citep{Kennicutt_2012_SFR}. We measured the fractional contributions of clumps~A, B, and C to the total flux—defined as the flux within a central 0.6\arcsec aperture—in both the UV continuum band (F475W) and the \OII\ map. The three clumps exhibit substantial differences in their relative flux contributions between the UV continuum band and the \OII\ map (see Table~\ref{tab:fesc}). Considering if the three clumps share similar dust attenuation and stellar populations, their fractional \OII\ fluxes should be consistent with their UV continuum fractions. The discrepancies for clumps~B and C, therefore, indicate that $f_{\mathrm{esc}} > 1$ likely arises from either incorrect dust attenuation estimates or a mismatch between their intrinsic stellar populations and the globally fitted model.

Therefore, to determine the spatial distribution of $f_{\mathrm{esc}}$, it is essential not only to identify individual LyC-emitting clumps but also to obtain high spatial resolution SED and dust maps.

It is worth noting that, to date, J1244-LyC1 and Haro~11 \citep{Komarova_2024} are the only systems that exhibit a clearly multi-clump spatial distribution of LyC leakage, and both are merger systems. 
Haro~11 is an extreme dwarf starburst galaxy hosting dozens of young massive clusters. 
Its two LyC-emitting knots show pronounced differences in their stellar populations. 
Although we cannot assert that J1244-LyC1 is a direct high-redshift analog of Haro~11, the similarities between the two systems are noteworthy. 
This may suggest the existence of a LyC photon leakage mechanism that does not strongly evolve with redshift.

\begin{deluxetable*}{lccccc}
\tablecaption{Spatially Resolved $f_{\mathrm{esc}}$ Calculation of J1244-LyC1 \label{tab:fesc}}
\tablehead{
\colhead{Target} &
\colhead{$\mathrm{mag}_{F225W}$} &
\colhead{$\mathrm{mag}_{F475W}$} &
\colhead{UV fraction\tablenotemark{a}} &
\colhead{[O\,II] fraction\tablenotemark{a}} &
\colhead{$(F_{F475W}/F_{F225W})_{\mathrm{obs}}$}
}
\startdata
J1244-LyC1 &  $27.81^{+0.20}_{-0.17}$ & $24.27^{+0.01}_{-0.01}$ & 1 & 1 &  $26.2^{+5.3}_{-3.8}$ \\
Clump-A   & $29.14^{+0.28}_{-0.23}$ & $26.87^{+0.05}_{-0.23}$ & $13.7\pm0.5\%$ & $6.2\pm2.8\%$  & $8.3^{+2.6}_{-1.6}$ \\
Clump-B   & $29.27^{+0.34}_{-0.27}$ & $27.79^{+0.11}_{-0.10}$ & $5.8\pm0.5\%$  & $5.5\pm2.5\%$  & $4.0^{+1.5}_{-0.9}$ \\
Clump-C   & $29.44^{+0.37}_{-0.28}$ & $28.47^{+0.21}_{-0.17}$ & $3.1\pm0.5\%$  & $10.5\pm2.8\%$ & $2.4^{+1.1}_{-0.6}$ \\
\enddata
\tablecomments{
All reported uncertainties represent 1$\sigma$ errors.
}
\tablenotetext{a}{Fraction of the total photometric flux of J1244-LyC1 measured within a 0.6\arcsec\ aperture.}

\end{deluxetable*}

\subsection{Can mergers boost LyC escape?}\label{subsect: merge_fesc}

Galaxy mergers as a mechanism for promoting LyC photon escape have long been actively discussed \citep{Jaskot_lowz_review_2025, Le_Reste_2025_III, zhu_lyman_2024, yuan_merging_2024,kostyuk_influence_2024}. At low redshift, the LzLCS survey constructed a sample of LyC leakers \citep{Flury_2022_I}, among which approximately 41\% exhibit merger signatures \citep{Le_Reste_2025_III}. During the cosmic noon epoch, mergers also appear to constitute the majority of LyC leakers \citep{zhu_lyman_2024, yuan_merging_2024, Wang_shengzhe_2026,Rivera-Thorsen_2026}, while cosmological simulations suggest that in the EoR, mergers can significantly enhance the reionization process \citep{kostyuk_influence_2024}.

However, how galaxy mergers promote LyC photon escape remains an open question. On one hand, mergers drive the inflow of low-metallicity cold gas into the central regions, possibly triggering central star formation and producing more LyC photons. Stellar feedback could then open low-density escape channels \citep{puskás_2025, Faria_2025,garaysolis_2025,Cenci_2024_merger_SB}. On the other hand, mergers strongly disturb the gaseous environment of galaxies, decreasing the covering fraction of neutral gas while making the ISM more turbulent \citep{garaysolis_2025, Purkayastha_2022_GPs, puskás_2025}. However, as \cite{kostyuk_influence_2024} pointed out, no comprehensive model has yet been established to fully describe this process, as none is currently capable of resolving all relevant scales, including those down to individual molecular clouds. More observational constraints are therefore required to understand whether merger systems promote global LyC escape or whether escape occurs only from specific substructures.

The diffuse LyC-band emission observed in J1244-LyC1 provides a unique opportunity to investigate this issue in the high-redshift Universe. We observe spatial variations of LyC photon leakage across different regions of a merger system at cosmic noon. As discussed in Section \ref{subsect: S_fesc_c}, we identify three LyC-leaking clumps in J1244-LyC1. Clump~A lies between the two UV-bright centers, while clumps~B and C exhibit spatial offsets from the main merger system.

For clumps~B and~C, the leakage positions are offset by less than 0.5\arcsec from the main body, similar to other merger-like LyC leakers \citep{gupta_mosel_2024, yuan_merging_2024}. Given their coincidence with the tidal tails, we attribute their LyC photon leakage to star formation occurring within these tidal features. For clump~A, however, its position between the two UV-bright centers makes its formation mechanism more uncertain. It is unclear whether the central leakage region belongs to one of the two merging galaxies, lies between them, or is located on the near side of the system. Regardless of the configuration, the presence of clump~A distinguishes J1244-LyC1 from other merger-featured LyC candidates: its LyC emission shows no significant spatial offset from the UV morphology. This confirms that a merger system can indeed produce strong LyC photon escape.

\subsection{The Formation of Massive LyC Leakers} \label{subsect: Massive}

A key question is whether the physical pathway enabling LyC escape in J1244-LyC1 differs substantially from that in other known LyC leakers. To date, confirmed LyC leakers at high redshift are predominantly low-mass systems $M_\ast \lesssim 10^{9.7}\,M_\odot$ with little dust attenuation $E(B-V) \lesssim 0.1$; e.g., \citealt{Fletcher_2019,yuan_merging_2024,gupta_mosel_2024,Ji_2025,Shapley_2016,marques-chaves_witnessing_2024,Me_tri__2025,kim2023smallregionbigimpact,Rivera_Thorsen_2022,Kerutt_2024,Wang_shengzhe_2026,Rivera-Thorsen_2026}). Massive LyC leakers are also rare in the low-redshift Universe, with roughly half of the known leakers having stellar masses below $10^{9}\,M_\odot$ \citep{Flury_2022_I,Le_Reste_2025_I,Le_Reste_2025_III,Jaskot_lowz_review_2025}. J1244-LyC1 is therefore unusual in being both massive and globally dusty $M_\ast=10^{10.2}\,M_\odot$; $E(B-V)=0.21$ while exhibiting a high $f_{\rm esc}=36\%\pm6\%$. Its high mass and SFR are comparable to the highest values seen in extreme emission-line galaxies at z$\sim 0.5$ \citep{Li2018}.

The massive low-redshift starbursts studied by \citet{Roy_2025} provide a particularly relevant comparison. Like J1244-LyC1, these systems are massive, dusty, and intensely star-forming. However, the three LyC detections in their five-object sample have low $f_{\rm esc}\lesssim3\%$, despite H\,\textsc{i}-only escape fractions of $33\%$--$84\%$ after correcting for dust attenuation. \citet{Roy_2025} interpreted this contrast as evidence that feedback-driven winds can produce low-H\,\textsc{i}-column sightlines in massive starbursts, while dust continues to absorb most of the potentially escaping LyC radiation. The substantially higher $f_{\rm esc}$ of J1244-LyC1 therefore cannot be explained simply by its global dust content. Instead, the difference may reflect the spatial distributions and effective covering fractions of neutral gas and dust along the LyC-emitting sightlines.

In low-mass, compact galaxies, stellar feedback can efficiently perturb the ISM and create low-density escape channels because of their relatively shallow gravitational potentials \citep[e.g.,][]{Rey_2022,Trebitsch_2017,Le_Reste_2025_III}. In massive galaxies, their larger gas reservoirs and deeper gravitational potentials make such restructuring more difficult \citep{Pandya_2021,Somerville_2015}, although sufficiently compact and intense starbursts may still drive winds capable of opening H\,\textsc{i}-transparent channels \citep{Roy_2025}. Mergers may provide an additional mechanism by both triggering intense star formation and redistributing the gas and dust on larger spatial scales. In J1244-LyC1, the disturbed morphology and spatially resolved LyC emission, together with its high total escape fraction, suggest that the merger may have created sightlines with low effective covering fractions of both H\,\textsc{i} and dust. This interpretation remains tentative because the dust distribution along the LyC escape paths is not directly constrained, but it offers a plausible explanation for the strong contrast with the massive low-redshift leakers.

The rarity of massive systems with high $f_{\rm esc}$ implies that the contribution of UV-luminous galaxies to the ionizing photon budget cannot be inferred from their abundance and intrinsic ionizing output alone, but depends critically on the physical processes that regulate $f_{\rm esc}$. J1244-LyC1 therefore provides a valuable opportunity to investigate how mergers may facilitate LyC escape from massive, dusty galaxies and to assess the possible contribution of such systems during the EoR.

\subsection{Environmental Effects} \label{subsect: environment}

% The impact of the environment on LyC leakage remains an open question.
% On one hand, an overdense environment may enhance star formation at $z>1$ \citep{Elbaz_2007, Taamoli_2024}; on the other hand, it remains unclear whether such environments facilitate the further escape of LyC photons into the IGM, or how this process evolves with redshift.
% xxx that paragraph did not say much new XXX
The impact of environment on LyC leakage remains an open question and may evolve strongly with redshift.
For searches targeting LyC leakers, the latter question is particularly relevant.  
It is commonly assumed that in the high-redshift Universe ($z>6$), overdense regions—such as protocluster environments—promote LyC photon escape into the IGM, and that Ly$\alpha$ emitters can trace the associated ionized structures.  
However, at $z<5.5$, some studies suggest that protocluster environments may be more neutral than the field \citep{Kashino_2025, Mawatari_2017, Liang_2021}.  
This could result from the formation of a circumgalactic medium (CGM) or from continued inflow of cold gas along large-scale structure, though no consensus has been reached.

% In the context of LyC searches, our high-redshift observational window is concentrated around the cosmic-noon epoch \citep{inoue_updated_2014, naidu_synchrony_2022}.  
% Many HST programs have targeted protocluster fields during this period, such as SSA22 \citep{Fletcher_2019} and the BOSS1244 field analyzed in this work.  
% Notably, the strongest LyC leaker in BOSS1244—J1244-LyC1—originates from a background field galaxy rather than from the protocluster itself.  

Against this background, the detection of J1244-LyC1 becomes particularly intriguing. 
Although it does not reside within the BOSS1244 protocluster (at $z \sim 2.24$), it is still affected by the IGM environment associated with the foreground protocluster. 
While we cannot definitively confirm that the line of sight toward J1244-LyC1 corresponds to a lower-than-average IGM density---as this is directly connected to the measured $f_{\mathrm{esc}}$---we can state with confidence that the presence of such large-scale protocluster structures does not entirely prohibit the detection of background LyC leakers.

Furthermore, there are no foreground galaxies within $\sim$3\arcsec\ of J1244-LyC1.  
This suggests that it is not strongly affected by the CGM of foreground systems—one reason it remains detectable. 

This serendipitous finding prompts reconsideration of whether systematic searches for LyC leakers should focus exclusively on protocluster fields. 
The presence of background LyC leakers may, in fact, provide additional information on the IGM along these lines of sight.

% Furthermore, there are no foreground galaxies within $\sim$3\arcsec\ of J1244-LyC1.  
% This suggests that it is not strongly affected by the CGM of foreground systems—one reason it remains detectable.  
% Future studies should examine more carefully whether the CGM of foreground galaxies impacts the detectability of LyC leakers and develop approaches to mitigate the observational biases introduced by such effects.

\section{Summary} \label{sect: Summary }

This work presents the first results from the MAMMOTH-LyC ultra-deep HST WFC3/UVIS imaging survey (HST-GO-17159; P.I.: X. Wang), targeting the core regions of two massive galaxy protoclusters at $z\sim2.2$. We discover a new strong LyC leaker at $z=2.39$, named J1244-LyC1, where spatially resolved LyC emission is detected using the ultra-deep F225W imaging acquired by MAMMOTH-LyC. The total LyC-band signal is detected at a $6\sigma$ level, with an estimated escape fraction of $f_{\mathrm{esc}} \sim 36\%$, and shows no spatial offset relative to the UV-band imaging (F475W). The galaxy is part of the EELG sample in the MAMMOTH-Grism survey \citep{Xin_Wang_2022_MAMMOTH}, and has been observed with HST WFC3/G141 grism spectroscopy, covering \OII and H$\beta$. In addition, two Keck/MOSFIRE $K$ band spectra cover its H$\alpha$ emission line. These observations robustly confirm the spectroscopic redshift of J1244-LyC1.
Within a 3\arcsec radius of J1244-LyC1, no foreground galaxies are found, and no unidentified strong emission lines are detected across 1.1--1.7~$\mu$m and 2.0--2.3~$\mu$m, confirming the absence of any foreground interloper.

% High-resolution imaging across multiple bands reveals that J1244-LyC1 exhibits  merger signatures---two photometric centers separated by a projected distance of 2.5~kpc and possible tidal-tail features. Thanks to the two Keck/MOSFIRE $K$ band observations, whose slits were placed at slightly different orientations, we detect two H$\alpha$ kinematic components separated by $\sim116\ {\rm km\,s^{-1}}$, which most likely trace the line-of-sight velocity difference between the two photometric components. This confirms that J1244-LyC1 is a major merger in its late stage. Since the global LyC emission shows no spatial offset, this represents the first high-$z$ merger system in which LyC photon escape has been directly confirmed.

High-resolution imaging across multiple bands reveals strong merger signatures in J1244-LyC1, including two photometric centers separated by a projected distance of 2.5~kpc and possible tidal features. 
The two Keck/MOSFIRE observations provide evidence for two H$\alpha$ kinematic components, with an adopted characteristic velocity separation of approximately $120~\mathrm{km\,s^{-1}}$, which most likely trace the line-of-sight velocity difference between the two photometric components. Together, these observations support a morphologically late-stage major-merger interpretation. The spatially resolved LyC emission demonstrates ionizing-photon escape from a high-redshift system with strong merger signatures.

We performed multi-band photometric SED fitting for J1244-LyC1 and verified the consistency between the derived physical parameters and the observed emission-line constraints. Because J1244-LyC1 exhibits spatially resolved LyC emission, we are, for the first time, able to study the substructures and physical processes responsible for LyC escape in a high-$z$ sample. Both the photometric centers and the tidal tails show evident LyC leakage. We interpret this as the result of vigorous star formation triggered by the merger process and the strongly disturbed ISM environment, jointly facilitating LyC photon escape. 
J1244-LyC1 represents an even rarer case of a massive LyC leaker. We further discuss the critical role of the merger process in enabling LyC escape in massive galaxies---unlike low-mass systems, massive galaxies rely more heavily on the strong ISM disturbances induced by mergers to open low-opacity channels for LyC leakage.

The dynamical structure, Ly$\alpha$ line profile, and spatial distribution of Ly$\alpha$ emission in J1244-LyC1 are therefore of great importance. Follow-up observations with HST and ground-based adaptive-optics IFUs, such as Keck/OSIRIS, will be crucial for further understanding. J1244-LyC1 represents a rare and ideal case for studying the impact of mergers on LyC leakage, providing valuable observational constraints on the role of merger systems in the reionization era.

\begin{acknowledgments}
We thank the anonymous referee for very constructive comments that helped improve the quality of this paper.
This work is supported by the National Natural Science Foundation of China (grant 12373009), the National Key R\&D Program of China No.2025YFF0510603, the CAS Project for Young Scientists in Basic Research Grant No. YSBR-062, the China Manned Space Program with grant no. CMS-CSST-2025-A06, and the Fundamental Research Funds for the Central Universities. XW acknowledges the support by the Xiaomi Young Talents Program, and the work carried out, in part, at the Swinburne University of Technology, sponsored by the ACAMAR visiting fellowship.
This work is also supported by NASA through HST grants HST-GO-16276 and HST-GO-17159.
This work is partly based on data obtained through Swinburne Keck program 2025A\_W335. RLD is supported by the Australian Research Council through the Discovery Early Career Researcher Award (DECRA) Fellowship DE240100136 funded by the Australian Government.
We thank those of Hawaiian ancestry on whose sacred mountain we are privileged to be guests. Without their generous hospitality, most of the observations presented herein would not have been possible.

\end{acknowledgments}

\bibliography{LyC_leaker_observation}
% \bibliographystyle{apj}

%
%%%%%%    %%%%%%%%%%%%%%%%%%%%%%%%%%%%%%%%%%%%%%%%%%%%%%%%%%%
%

\appendix \label{sect: app}

\section{ASTRODRIZZLE INPUT PARAMETERS} \label{sect: app_B}

Our input parameters for {\tt Astrodrizzle} are listed in Table ~\ref{tab:drizzle}.

\begin{table*}[ht] 
\centering
\caption{AstroDrizzle Parameters}
\label{tab:drizzle}
\begin{tabular}{lll}
\hline\hline
\textbf{Parameter} & \textbf{Value} & \textbf{Description} \\
\hline
% \texttt{output} & \texttt{iexi1244\_f225w\_30ms\_17\_iter9\_new} & Output drizzled image name \\
% \texttt{build} & True & Generate all intermediate products \\
% \texttt{driz\_sep\_refimage} & \texttt{''} & No reference image for separate drizzle \\
% \texttt{num\_cores} & 30 & Number of CPU cores used \\
% \texttt{restore} & False & Do not restore previous products \\
% \texttt{clean} & True & Remove temporary files \\
% \texttt{preserve} & False & Do not preserve intermediate outputs \\
\texttt{skysub} & True & Perform sky subtraction \\
\texttt{skymethod} & globalmin+match & Sky estimation method \\
% \texttt{driz\_separate} & True & Create single-image drizzle products \\
% \texttt{driz\_sep\_wcs} & True & Use WCS for separate drizzle \\
% \texttt{driz\_sep\_ra} & \texttt{Ra} & Output RA center for separate drizzle \\
% \texttt{driz\_sep\_dec} & \texttt{Dec} & Output Dec center for separate drizzle \\
% \texttt{driz\_sep\_crpix1} & \texttt{X} & CRPIX1 for separate drizzle \\
% \texttt{driz\_sep\_crpix2} & \texttt{Y} & CRPIX2 for separate drizzle \\
% \texttt{driz\_sep\_outnx} & \texttt{outnx} & Output size in X \\
% \texttt{driz\_sep\_outny} & \texttt{outny} & Output size in Y \\
% \texttt{driz\_sep\_rot} & \texttt{rot} & Position angle of output frame \\
\texttt{driz\_sep\_scale} & 0.03 & Pixel scale for separate drizzle (arcsec/pixel) \\
% \texttt{median} & True & Create median image \\
% \texttt{median\_newmasks} & True & Generate new masks for median combination \\
\texttt{combine\_type} & imedian & Final combine method \\
\texttt{combine\_nsigma} & 4 3 & Low/high sigma clipping thresholds \\
% \texttt{combine\_nlow} & 0 & Number of low pixels to reject \\
% \texttt{combine\_nhigh} & 0 & Number of high pixels to reject \\
% \texttt{combine\_lthresh} & None & No low threshold \\
% \texttt{combine\_hthresh} & None & No high threshold \\
\texttt{combine\_grow} & 1 & Pixel grow radius for rejection \\
% \texttt{blot} & True & Blot median image back to input frame \\
\texttt{driz\_cr} & True & Cosmic-ray rejection enabled \\
\texttt{driz\_cr\_corr} & True & Cosmic-ray correction enabled \\
% \texttt{driz\_combine} & True & Create final drizzle-combined product \\
\texttt{final\_wht\_type} & IVM & Inverse-variance weighting \\
\texttt{final\_kernel} & square & Drizzle kernel \\
\texttt{final\_pixfrac} & 0.8 & Pixel fraction for final drizzle \\
% \texttt{final\_fillval} & None & No fill value applied \\
% \texttt{final\_wcs} & True & Use WCS for final drizzle \\
% \texttt{final\_refimage} & \texttt{''} & No final reference image \\
% \texttt{final\_ra} & \texttt{Ra} & Final output RA center \\
% \texttt{final\_dec} & \texttt{Dec} & Final output Dec center \\
% \texttt{final\_crpix1} & \texttt{X} & Final CRPIX1 \\
% \texttt{final\_crpix2} & \texttt{Y} & Final CRPIX2 \\
% \texttt{final\_outnx} & \texttt{outnx} & Final output X size \\
% \texttt{final\_outny} & \texttt{outny} & Final output Y size \\
% \texttt{final\_rot} & \texttt{rot} & Final rotation angle \\
% \texttt{final\_scale} & 0.03 & Final pixel scale (arcsec/pixel) \\
% \texttt{runfile} & \texttt{iexi1244\_total\_30ms.log} & Log file name \\
\hline
\end{tabular}
\end{table*}

\section{Keck Data Reduction} \label{sect: app_C}
% still in progress
% \textcolor{red}{Still in progress}

The reduction of Keck MOSFIRE data was performed using the PypeIt pipeline \citep{Prochaska_pypeit_2020}. The standard processing sequence included flat-fielding, dark subtraction, cosmic ray detection and slit tracing. Wavelength calibration was derived directly from OH sky lines present in the science frames. One-dimensional spectra were subsequently generated using the Optimal Extraction algorithm. Regarding object identification, the pipeline is configured by default to extract objects in the slit center. While this setup was successful for most sources in \cite{Zhou_hang_2025}, J1244-LyC1 in the first observations was offset from the slit center, causing the default automatic tracing algorithm to fail. Consequently, we lowered the required signal-to-noise ratio threshold to successfully trace and extract this specific source. Flux calibration was achieved using standard star observations processed identically to the science targets. A specific modification was required for the second set of observations, which utilized the long-slit mode. The default PypeIt configuration treats long-slit data as star traces; this misinterpretation leads to failures in source extraction. To resolve this, we customized the slit parameters by manually adding the slit traces to force extraction, which enabled PypeIt to correctly identify and extract the source.

\section{Seeing Convolution and MOSFIRE Slit Placement}
\label{app:slit_test}

\begin{figure*}[htbp]
\centering
\includegraphics[width=\textwidth]{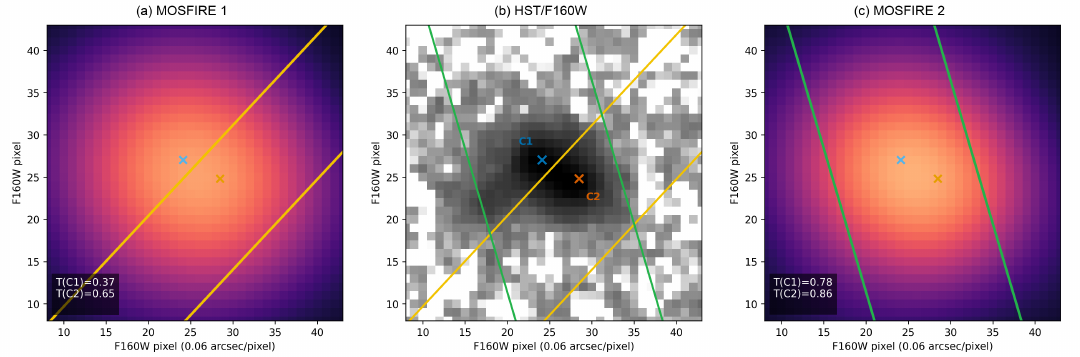}
\caption{Seeing-convolution and slit-placement test for J1244-LyC1. (a) F160W \texttt{GALFIT} model convolved with a Gaussian PSF of FWHM $0.7\arcsec$, with the $0.7\arcsec$ slit of the first observation overlaid. The estimated throughputs are $T_1({\rm C1})\simeq0.37$ and $T_1({\rm C2})\simeq0.65$. (b) HST/F160W image with the slit boundaries of the first and second observations shown in yellow and green, respectively. Blue and orange crosses mark the centroids of C1 and C2. (c) The corresponding model for the second observation, with a seeing FWHM of $0.6\arcsec$ and a slit width of $1.0\arcsec$, yielding $T_2({\rm C1})\simeq0.78$ and $T_2({\rm C2})\simeq0.86$. These throughputs describe the sampling of the F160W model light and should not be interpreted directly as H$\alpha$ line-flux fractions.}
\label{fig:slit_test}
\end{figure*}

To assess the effects of seeing and slit placement, we convolved the separate F160W \texttt{GALFIT} models of C1 and C2 with Gaussian PSFs corresponding to the seeing of each MOSFIRE observation and integrated the resulting light within the respective slit footprints (Figure~\ref{fig:slit_test}). The component throughputs denote the fractions of each component's own F160W model flux entering the slit. Both components contribute to the first observation after seeing convolution, while the second observation admits a larger fraction of both components, particularly C1.

The increased sampling of C1 is qualitatively consistent with the additional H$\alpha$ velocity component becoming more apparent in the second spectrum. However, the H$\alpha$ emission need not follow the F160W morphology, and the model does not include possible diffuse or tidal line emission. These estimates therefore illustrate the changing slit weighting but do not establish a unique correspondence between the photometric and kinematic components.

\section{spectrum} \label{sect: app_D}
\begin{figure*}[htbp]
    \centering
    \includegraphics[width=1\textwidth]{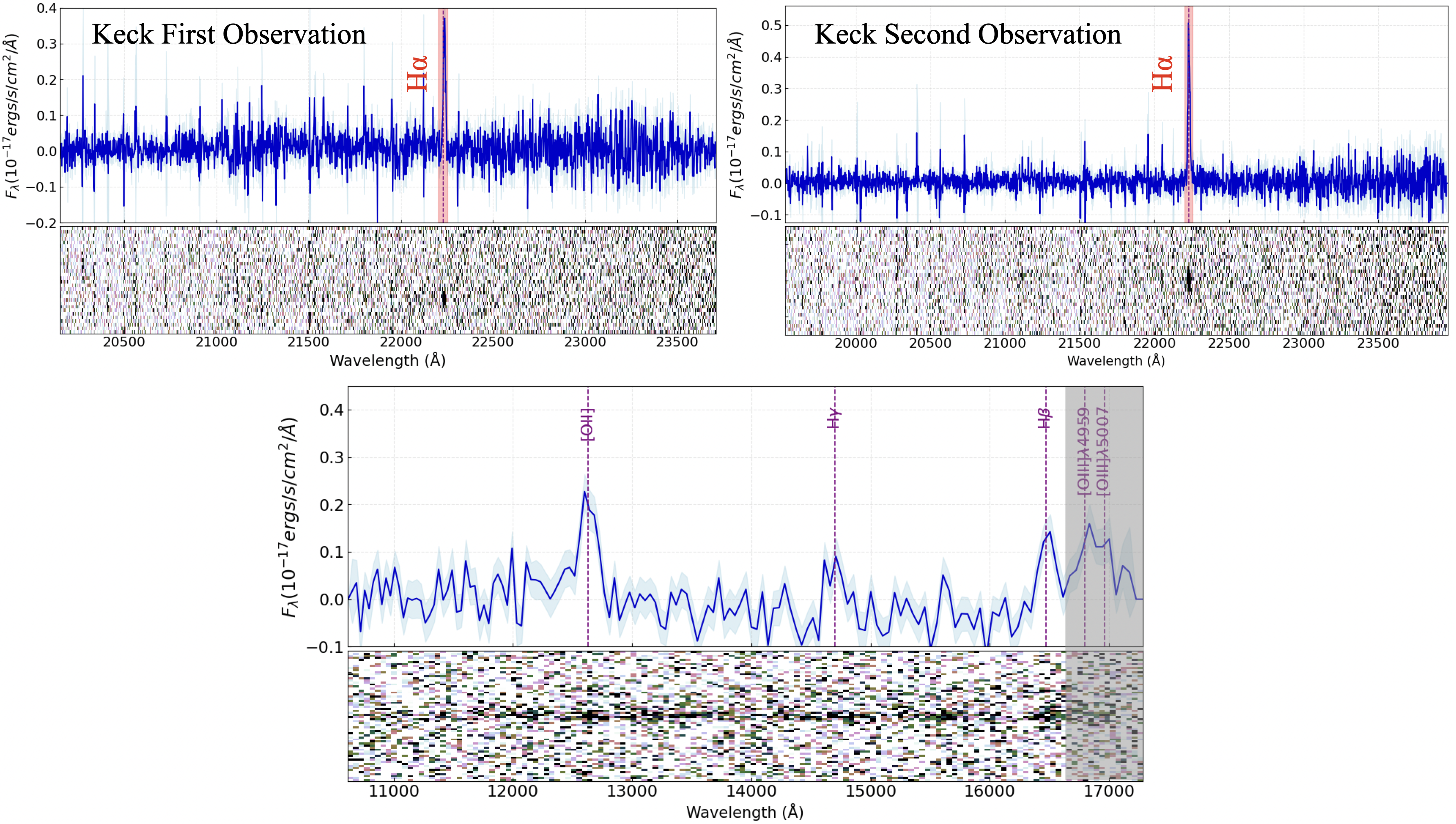}
    % \parbox[b]{6.9in}{\footnotesize {\bf Fig.~1.}
    \caption{
    Spectroscopic observations of J1244-LyC1. {\sc Top left }and {\sc top right panels}: Keck MOSFIRE $K$ band spectra. {\sc Bottom panel}: HST WFC3/G141 grism spectrum, with the region affected by the G141 edge indicated by the gray shaded area.
    }
    \label{app_fig:1}
\end{figure*}

%%%%%%%%%%%%%%%%%%%%%%%%%%%%%%%%%%%%%%%%%%%
\end{document}